\documentclass[10pt,a4paper]{article}

\usepackage[left=2cm,right=2.5cm, top=2cm,bottom=3cm]{geometry}
\usepackage[utf8]{inputenc}
\usepackage[english,russian]{babel}
\usepackage[pdftex]{graphicx}
\usepackage{amsmath}
\usepackage{amsthm}
\usepackage{amssymb}
\usepackage{amsfonts}
\usepackage{eucal}
\usepackage{setspace}
\usepackage{mathtools}
\usepackage{afterpage}
\usepackage[svgnames]{xcolor}
\usepackage{cancel}
\usepackage[breaklinks,colorlinks,allcolors=DarkCyan,unicode,pdftitle={Universal asymptotics}]{hyperref}

\usepackage{authblk}
\usepackage{times}

\usepackage[sort&compress]{natbib}

\theoremstyle{remark}
\newtheorem{remark}{Remark}

\providecommand{\keywords}[1]
{{
  \small	
  \textbf{\itshape{Keywords ---}} #1
}}

\numberwithin{equation}{section}
\numberwithin{remark}{section}

\renewcommand{\=}{\stackrel{\mbox{\scriptsize def}}{=}}
\newcommand{\I}{\text{i}}
\renewcommand{\d}{\text{d}}
\renewcommand{\v}{v}
\renewcommand{\O}{\Omega}

\renewcommand{\t}{\tau}
\newcommand{\T}{T}
\newcommand{\TF}{\mathcal{T}}
\newcommand{\WF}{\mathcal{W}}
\newcommand{\UF}{\mathcal{U}}
\newcommand{\X}{X}
\newcommand{\Co}{\mathcal C}
\renewcommand{\l}{\ell}

\newcommand{\e}{\text{e}}
\newcommand{\Exp}[1]{\operatorname{exp}\left( #1\right)}
\newcommand{\const}{\text{const}}

\newcommand{\F}[1]{\mathcal{F}\{#1\}}

\newcommand{\abs}[1]{\left| #1 \right|}

\def\cc{\mathrm{c.c.}}

\newcommand\GA{\varGamma}

\newcommand\pd[2]{\frac{\partial#1}{\partial#2}}
\newcommand\pdd[2]{\frac{\partial^2#1}{\partial#2^2}}
\newcommand\ppdd[3]{\frac{\partial^2#1}{\partial#2\,\partial#3}}
\newcommand\dd[2]{\frac{\d #1}{\d#2}}
\newcommand\bcdot{{\boldsymbol\cdot}}

\def\SQRTO{\sqrt{1-\Omega_0^2}}

\newcommand\tp{\hat T}	
\newcommand\ttp{\hat t}	

\newcommand{\p}{\mathcal P}
\newcommand{\pp}{\boldsymbol{\mathcal P}}
\newcommand{\ppp}{\tilde{\boldsymbol{\mathcal P}}}
\newcommand{\J}{\mathcal J}
\newcommand{\LL}{\mathbb L}
\newcommand{\bPsi}{\boldsymbol\Psi}

\begin{document}

\selectlanguage{english}

\title{Universal 
formal 
asymptotics for localized oscillation 
of a discrete
mass-spring-damper system of time-varying properties, embedded into a one-dimensional
medium described by the telegraph equation
with variable coefficients}

\author[1]{Serge N. Gavrilov}
\author[2]{Ilya O. Poroshin}
\author[1]{Ekaterina V. Shishkina}
\author[1]{Yulia A. Mochalova}
\affil[1]{Institute for Problems in Mechanical Engineering RAS, St.~Petersburg, Russia}
\affil[2]{Peter the Great St. Petersburg Polytechnic University (SPbPU), Russia}


\maketitle

\selectlanguage{english}

\begin{abstract}	
We consider a quite general problem concerning a linear free
oscillation of a discrete mass-spring-damper system. This discrete sub-system is embedded 
into a one-dimensional
continuum medium described by the linear
telegraph equation.  
In a particular case, the discrete sub-system can move along the continuum one
at a sub-critical speed.  
Provided that the dissipation in both discrete and continuum sub-systems is
absent, if parameters of the sub-systems are constants, under
certain conditions (the localization conditions),
a non-vanishing oscillation localized near the discrete
sub-system can be possible.
In the paper we assume that 
the dissipation in the damper and the medium is 
small, and
all discrete-continuum system parameters are
slowly varying functions in time and in space (when applicable), 
such that the localization condition 
is fulfilled for the instantaneous values of the parameters in a certain
neighbourhood of the discrete sub-system position. 
This general statement can describe a number of mechanical systems of various
nature.
We derive the expression for the
leading-order term of a universal asymptotics, 
which describes a localized oscillation 
of the discrete sub-system. In the non-dissipative case, the leading-order term of the expansion for the amplitude
is found in the form of an algebraic expression, which involves the instantaneous values of the system
parameters.
In the dissipative case, the leading-order term for the amplitude, generally, is found in
quadratures in the form of a functional, which depends on the history
of the system parameters, though in some exceptional cases the result can be
obtained as a function of time and the instantaneous limiting values of the system
parameters.
In previous studies, several non-dissipative particular cases of the
problem under consideration are investigated using a similar approach,
provided that only one parameter of the discrete-continuum system is 
a slowly time-varying 
non-constant quantity, whereas all other parameters are constants.
We show
that asymptotics obtained in previous studies are particular cases of the
universal asymptotics. The existence of a universal asymptotics in the form of
a function is a non-trivial fact, which does not follow from the summarization
of the 
previously obtained results. 
Finally, we have justified the universal asymptotics by
numerical calculations for some particular cases.
\end{abstract}


\section{Introduction}
Consider a linear free
oscillation of a discrete mass-spring system. The discrete system is embedded 
into a one-dimensional
continuum medium, described by the linear
Klein-Gordon partial differential equation (PDE). Such a coupled discrete-continuum system is very similar 
to the one considered in
classical Lamb study \cite{Lamb1900}, where the wave equation was used instead
of the Klein-Gordon one. Lamb showed that embedding a linear mass-spring system
into the medium leads to transmission of energy outside from the discrete
sub-system by running waves. As a result, free oscillation of the mass-spring 
system vanishes in the same way as it happens, when a damper is
added into an isolated mass-spring system. 
In our case, when the medium is described by the
Klein-Gordon equation, the dynamics of the discrete-continuum system essentially depends 
on the system parameters. 
Under certain conditions (the localization conditions) fulfilled in some domain in the
problem parameter space (the localization domain), a free
oscillation of the mass-spring system remains non-vanishing, as we
observe in the case of an isolated discrete mass-spring system.
This happens due
to the existence of a trapped mode
\cite{Abramyan1994,Glushkov2011a,mciver2003excitation,Mishuris2020,pagneux2013trapped,porter2007trapped,porter2014trapped,ursell1951trapping,Ursell1987,Ind-book-R2E,Kuznetsov2002,kaplunov1995simple,Voo2008,Indeitsev2006},
to which oscillation, spatially localized near the discrete sub-system,
corresponds. The trapped mode frequency belongs to the stop-band of dispersion
characteristics for the continuum system; thus we deal with so-called strong
localization \cite{luongo2001mode,Luongo1992}.
For the discrete sub-system
moving along the continuum one at a constant sub-critical speed,  the
trapped mode also can exist \cite{gavrilov2002etm,Gavrilov2022jsv}. 
If the localization conditions are not fulfilled, then we observe a power-law
vanishing oscillation of the mass-spring system \cite{Shishkina2023jsv}. 

In the present paper, we extend the discussed above discrete-continuum (or
composite) system in several aspects:

\begin{enumerate}       
\item A small viscous friction in the continuum sub-system is taken into
account;
thus, the Klein-Gordon PDE transforms into the telegraph PDE
\cite{MyintU2007};
\item A small  viscous friction in the discrete sub-system is taken into
account;
thus, we deal with a mass-spring-damper system;
\item Parameters of the discrete mass-spring-damper system 
are slowly time-varying functions;
\item Parameters of the continuum system 
are slowly varying functions of time and the spatial co-ordinate;
\item In a particular case, the discrete sub-system can move along the
continuum one at a sub-critical speed\footnote{The absolute value of the load
speed is less than the speed of sound}, which is also
assumed to be a slowly time-varying function.

\end{enumerate}

Thus, the problem for the composite system with constant parameters we have
discussed above at the beginning of the Introduction is the zeroth order problem
for such an extended problem formulation.
All necessary results for this zeroth order problem are obtained, e.g., in
\cite{Gavrilov2022jsv}.
We require that the localization conditions
are always fulfilled for the instantaneous values of the system parameters in a certain
neighbourhood of the discrete sub-system. Therefore, one can expect that the
motion of the discrete sub-system is close to the oscillation with the same 
frequency as it is observed for zeroth order problem (i.e., with an instantaneous value
of the trapped mode
frequency). 
The amplitude of the
oscillation is expected to be a slowly time-varying function.

The extended problem formulation suggested above can describe a number of
physical systems of a various nature. Most often, when considering similar
problems in the framework of the classical mechanics, see, e.g.,
\cite{abramian2017oscillations,Glushkov2011a,kaplunov1986vibrations,kaplunov1995simple,Ind-book-R2E,Roy2018,Kruse1998},
it is assumed that the motion of the continuum system 
corresponds to transverse oscillation of a string on an elastic foundation,
though 
longitudinal 
\cite{Shatskyi2021}
or rotational \cite{kaplunov1986torsional} oscillation of a rod can be under
investigation. The electric current oscillation in the transmission lines also is a
possible application \cite{Voo2008}. In the paper, we generally assume that we
deal with a mechanical system, but the results can be applied to other
physical systems. 
The possible applications can be related to ageing processes \cite{abramian2017oscillations}, design
of dynamic materials
\cite{Blekhman2000,Lurie2007,Rousseau2010},
ice-induced vibrations \cite{Abramian2019}, rain-wind
vibrations \cite{Zhou2020}.
The
problems, which involve the motion of the discrete sub-system along the
continuum one, are the classical moving loads problems
\cite{fryba1972vibration}, which have engineering applications, e.g., related to 
railways \cite{Roy2021}. 
Note that the model of a
string on the Winkler foundation is frequently used in mechanics of systems
with moving loads to obtain the
test analytic solutions
\cite{fryba1972vibration,Roy2018,kaplunov1986vibrations,Gil2020,DasGupta2023,Kruse1998}. 

The simplest particular case of a system with time-varying parameters is an
isolated mass-spring system with a time-varying stiffness. The behaviour of
this system can be approximately described by WKB approach, which is initially based on
studies by Carlini \cite{Carlini1818}, 
Liouville \cite{Liouville1837} and 
Green \cite{Green1838,Green2014}. It was demonstrated that in the first
approximation, the amplitude is
proportional to the inverse of the square root of the natural frequency.
%
This formula now is well-known as the 
Liouville--Green approximation, later 
rediscovered by Rayleigh \cite{Rayleigh1912}, and
as (J)WKB approximation after studies
\cite{Jeffreys1925,Wentzel1926,Kramers_1926,Brillouin1926}. The historical
aspects are discussed in 
\cite{feschenko1967eng,McHugh1971,Froeman2002}.
In Nayfeh book \cite{nayfeh2008perturbation},
a formal asymptotic procedure based
on the general ideas of the method of multiple scales, which
allows one to obtain such a result in a simple manner, is suggested.
For the reader's
convenience, we consider a discrete mass-spring-damper system with time-varying
properties using the Nayfeh multiple scales approach in Appendix~\ref{App-1d}. This simple
example helps us to introduce and discuss some peculiarities of the problem, 
which play an important role when considering the composite discrete-continuum 
system with variable properties.

The similar asymptotic approach in spirit of the
method of multiple scales 
was first time applied to describe non-stationary localized
oscillation in a coupled discrete-continuum system
with time-varying properties in \cite{gavrilov2002etm}, where an oscillation of an
inertial load, non-uniformly moving along an infinite string on the Winkler
foundation, was considered. The speed of the load was assumed to be a slowly
time-varying function.
Though the model of a string on the Winkler foundation is widely used in engineering,
and such a problem has many important applications, before \cite{gavrilov2002etm} 
there were no
approximate solution of a simple structure.
Indeed, this is an essentially non-stationary problem, which takes into account
both an inertial character of the load \cite{kaplunov1986torsional} and a
non-uniform regime of motion \cite{kaplunov1986vibrations,Gavrilov1999jsv}.
The method proposed in \cite{gavrilov2002etm} can be treated as an extension
of the Nayfeh formal procedure of WKB approach,
which deals with an ordinary differential equation (ODE), to
the case of a dispersive hyperbolic PDE with two independent variables coupled with ODE.
Now we have realized that the assumed representation for the solution 
in \cite{gavrilov2002etm}
(an asymptotic
ansatz) has the structure very similar to the one used in the framework 
of the space-time ray method
\cite{Babich2002,Babich2009}. 
The free oscillation was assumed in the form, which we call
{a single-mode ansatz}, corresponding to the frequency of a localized
oscillation.
To describe a non-stationary free oscillation in the discrete sub-system, a
problem for a PDE,
coupled with an ODE, 
in the first approximation was asymptotically reduced to 
a linear quite lengthy ODE
with variable coefficients
of the first order for the unknown 
amplitude. {This is an essential simplification compared with the general
space-time ray method procedure, which is related to the fact that we deal with
an oscillation 
localized  near the discrete sub-system
only, and do not try to
continue the solution outside.}
Finally, 
for an arbitrary law of
the load speed variation,
the oscillation amplitude was found in a form of
a simple algebraic expression, which depends on the instantaneous value of the
speed.
The expression obtained involves the unknown initial
amplitude and phase (or, equivalently, the complex amplitude). To find the
unknown constant, the obtained asymptotic expansion was matched 
with the expression obtained by
the method of stationary phase applied to the same system with
constant parameters equal to the corresponding initial values. The
obtained asymptotic solution was verified through the independent numerical
calculations, and an excellent agreement was shown.



In
\cite{Gavrilov2022jsv,Gavrilov2019nody,Gavrilov2019asm,indeitsev2016evolution,gavrilov2002etm},
the described above approach was applied to various particular cases of the
general system considered in this paper, as well as to a more complicated
problem where the continuum system was a Bernoulli-Euler beam
\cite{Shishkina2019jsv}. The resonant case, i.e., passage through a resonance,
was considered in 
\cite{Shishkina2020}.
In all these studies, the only one of the composite system
parameters was assumed to be a function depending on time, 
whereas all other parameters were taken as constants. Again, the final
expressions for the amplitude were found as the algebraic expressions,
which are valid for an
arbitrary law of
the parameter variation.
In particular, in 
\cite{Gavrilov2019nody} a localized oscillation
of a string with time-varying tension was considered.
In the last case, the continuum system is
described by PDE with time-varying coefficients, whereas in all other studies
the corresponding PDE can be reduced to a PDE with constant coefficients.
The asymptotics obtained in such a way is so-called formal asymptotics, i.e.
it satisfies the corresponding equations and initial conditions up to the
certain asymptotic accuracy. 
Moreover,
when constructing the asymptotics, a number of not well grounded 
assumptions were used. The most voluntaristic of them is the possibility to
represent a non-stationary oscillation of a system 
with time-varying parameters in the form of a single-mode ansatz, which
corresponds to ``a~pure localized oscillation'' at a given frequency. 
Dealing with PDE with time-varying coefficients, we do
not even have the orthogonality conditions, which allow one to separate
oscillations with different frequencies. Another question is the validity
of the matching procedure for two asymptotics, which were obtained by
entirely different approaches. 
Thus, in our opinion, the analytic results should be at least justified by
independent numerical calculations. This was done in the framework of
every problem for 
various regimes of parameters time-varying,
and various loadings.

The current paper generalizes the analytic work in series of papers 
\cite{Gavrilov2022jsv,Gavrilov2019nody,Gavrilov2019asm,indeitsev2016evolution,gavrilov2002etm}.
Compared to previous works, the novelty of this paper is assured by
the following results:
\begin{enumerate}     
\item
In previous studies, {\it only one} parameter of the composite system was assumed to
be a slowly time-varying quantity, whereas all other parameters
were constants. Now {\it each parameter is an independent slowly time-varying
function of time,
and additionally each parameter of the continuum system is a slowly varying
function of the spatial co-ordinate.} Nevertheless, in the
non-dissipative case, 
we still obtain the leading-order term of the expansion for the amplitude of
localized oscillation as {\it an algebraic expression, 
which involves the instantaneous limiting values of the system
parameters independently varying accordingly to unknown arbitrary laws.} 
To justify the applicability of the formal asymptotics, a particular case of
the composite system with two independently time-varying parameters is investigated
numerically, and an excellent agreement was demonstrated, see
Sect.~\ref{Sect-two-pa}.

\item 
In previous studies,  we did not introduce the dissipation.  Now, the
small viscous friction in both sub-systems is taken into account. In the
dissipative case, we generally obtain the leading-order term of the expansion for the
amplitude {\it in quadratures} as a functional, which depends on the history
of the system parameters, though in some exceptional cases the result can be
found as a function of time and the instantaneous limiting values of the system
parameters.
A particular case of the system with a time-varying parameter and a viscous
friction in a discrete sub-system is
investigated numerically, and an excellent agreement is demonstrated, see
Sect.~\ref{Sect-num-diss}.

\item
In the present paper, the corresponding asymptotics is obtained in the form of
a universal formula. This formula involves all previous results 
of studies 
\cite{Gavrilov2022jsv,Gavrilov2019nody,Gavrilov2019asm,indeitsev2016evolution}
as very particular cases, see Sect.~\ref{sect-previous}.
This result is not the summarization of previously
obtained results, but it is derived due to the mathematical trick firstly
suggested by Poroshin in \cite{Gavrilov2022jsv}. The trick is
related to the lucky choice of variables depending on the system
parameters, which allows us to guess how to represent, in the non-dissipative case, the
lengthy right-hand side 
of the first approximation
equation in the form of the exact derivative for a function of the variables,
see Sect.~\ref{sect-disc}.

\item We demonstrate that the excellent practical applicability of the 
constructed formal asymptotics
is related, in our opinion, with the phenomenon of
the anti-localization of non-stationary linear waves 
\cite{Gavrilov2024,Shishkina2023cmat,Shishkina2023jsv,Gavrilov2023DD}, see Sect.~\ref{As-breaks}.

\end{enumerate}

{To understand better the difference between the cases when the amplitude can
be found as a function of instantaneous values of the system parameters, or as a functional depending on the history of the
parameters, we recommend to the reader to look through the material of
Appendix~\ref{App-1d}.} The fact that 
the leading-order term for 
the amplitude can be found in
the form of a function
in the case of multiple
time-varying parameters 
is surprising for us. Apparently, from the mathematical
point of view, it follows from some unobvious properties of the governing
equations. On the other hand, from the engineer's perspective, 
the result in the form of a function, which is an algebraic expression,
gives a comprehensive understanding of the system behaviour, whereas
the result in the form of a functional
requires additional calculations to visualize it even for a given history of
the system
parameters, see, e.g.,
Sect.~\ref{Sect-two-pa}, \ref{Sect-num-diss}, respectively.



\section{Mathematical formulation}
\label{Sect-formulation}

Consider the following coupled system of equations:
\begin{gather}
\label{eq2}
\frac{\d }{\d t}
\left( M \frac{\d \UF}{\d t} \right)
{+2\epsilon\GA \dd{\UF}{t}}
+ K \UF = -P(t) +p(t),
\\
\label{eq1}
\frac{\partial }{\partial x}\left( \TF  \frac{\partial u}{\partial x} \right)
- \frac{\partial }{\partial t}
\left(
\rho \frac{\partial u}{\partial t} \right)
{- 2\epsilon\gamma
\pd{u}{t}}
- k u = -P(t)\delta(x-\l(t)).
\end{gather}
We assume that Eqs.~\eqref{eq2}, \eqref{eq1} describe the discrete and continuum  
sub-systems of the composite system under investigation,
respectively. Equation \eqref{eq2} is an ordinary differential equation, which,
clearly, describes a motion of a mass-spring-damper system. 
Here $t$ is time, $M$ is the mass, $K$ is the stiffness,
$\epsilon\varGamma$ is the damping. The quantity $\epsilon>0$ is the
dimensionless formal small parameter; thus, the damping is assumed to be
small.
The quantity
$p(t)$ in the right-hand side of Eq.~\eqref{eq2} is a given external force,
whereas $P(t)$ is an unknown internal interaction force between the
sub-systems. The assumed restrictions on function $p(t)$ are discussed 
in detail in what follows in this section. The parameters of the
discrete sub-system are assumed to be given smooth slowly time-varying
functions of $\epsilon t$:
\begin{equation}
K=K(\epsilon t),
\qquad
M=M(\epsilon t),
\qquad
\varGamma=\varGamma(\epsilon t),
\end{equation}
such that 
\begin{gather}
M\geq0.
\label{restr-par-11}
\end{gather}

The first term in Eq.~\eqref{eq2}, accounting effects related to a time-varying
mass, is written in the form, which assumes a mass supply into the discrete
sub-system accompanied by zero
momentum supply, see 
\cite{Mescherskiy1952,Irschik2004,LeviCivita1928,LeviCivita1928a,Zhilin2003}. 
Analogous formulation is used, e.g., in studies
\cite{abramyan2011oscillations,abramian2014oscillations}.
\begin{remark}	
\label{remark-mass}
Alternatively, one can assume that the term 
$2\epsilon\varGamma \frac{\d\mathcal U}{\d t}$
in the right-hand side of Eq.~\eqref{eq2}
is not related to a dissipation
itself, but 
the sum of two terms 
\begin{equation}
\frac {\d M}{\d t}\frac{\d\mathcal U}{\d t}+
2\epsilon\varGamma \frac{\d\mathcal U}{\d t}
\end{equation}
in Eq.~\eqref{eq2} represents 
a reactive force, which is proportional to the
particle velocity of the discrete sub-system,
see Eq.~(3.22) in \cite{Irschik2004},
\cite{Holl1999}.
In such a way, we can take
into account non-zero momentum supply into the discrete sub-system. In the
framework of such a physical interpretation, quantity $\varGamma$ is related to the
rate of the momentum supply. Accordingly, in the paper, 
we generally do not restrict ourselves with the case 
$\varGamma\geq0$, which corresponds to the presence of the viscous friction in
the discrete sub-system.
\end{remark}


In the same way, as in previous paper \cite{Gavrilov2022jsv},
we assume that the stiffness $K$ can be positive (stabilizing), {negative
(destabilizing)}, or zero:
\begin{equation}
       K\lesseqqgtr0. 
\end{equation}
The destabilizing effect from the discrete spring with a negative stiffness
can be compensated by stabilizing elastic foundation with $k>0$. The stability
condition for the coupled composite system is Eq.~\eqref{stab-cond}.
Discussing destabilizing springs is quite natural in the context of localized
oscillation, since the presence of a pure-spring inclusion with $K<0$ can lead
to linear waves localization, see again Eq.~\eqref{stab-cond}.
A destabilizing spring can be considered as
an oversimplified model of a crack, which can also cause the linear wave
localization, see \cite{Glushkov2006a,porter2014trapped}.
Note, that destabilizing springs are used when constructing meta-materials
\cite{chronopoulos2015enhancement,ahn2014active,huang2014vibration,li2013negative,oyelade2017dynamics,pasternak2014chains,wu2014analysis}.


Equation \eqref{eq1} can describe a number of mechanical systems of a various
nature. Most often, see, e.g.,
\cite{abramian2017oscillations,Glushkov2011a,kaplunov1986vibrations,Roy2018,Kruse1998,DasGupta2023,kaplunov1995simple,Ind-book-R2E} 
it is assumed that the motion of the continuum system 
corresponds to transverse oscillation of a string on an elastic foundation,
though 
longitudinal 
\cite{Shatskyi2021}
or rotational \cite{kaplunov1986torsional} oscillation of a rod can be under
investigation.
The corresponding
displacement is $u(x,t)$, where $x$ is a spatial co-ordinate.
The parameters of the continuum sub-system are assumed to  be given smooth,
slowly varying functions of $\epsilon x$ and $\epsilon t$:
\begin{equation}
\mathcal T=\mathcal T(\epsilon x, \epsilon t),
\qquad
\rho=\rho(\epsilon x, \epsilon t),
\qquad
k=k(\epsilon x, \epsilon t).
\qquad
\gamma=\gamma(\epsilon x, \epsilon t),
\label{all-slow}
\end{equation}
such that 
\begin{gather}
\mathcal T>0,
\qquad
\rho \geq0,
\qquad
k>0.
\label{restr-par-21}
\end{gather}
Here $k$ has the meaning of the elastic foundation stiffness, $\rho$ is
the mass density (or the moment of inertia) depending on the nature of physical processes
under consideration. Quantity $\mathcal T$ is the string tension or
the corresponding elastic
modulus (depending on the physical interpretation). 
Additionally, the damping in foundation per unit length $\epsilon\gamma$ is assumed to be a
small quantity.
Again, the second term in Eq.~\eqref{eq1} accounting effects related to a time-varying
mass per unit length $\rho(\epsilon x,\epsilon t)$, is formulated in the form,
which assumes supply of mass accompanied by zero supply of momentum, or the
reactive force proportional to the particle
velocity $\frac{\d u}{\d t}$ as discussed in Remark~\ref{remark-mass} is under
investigation.
In the paper, 
we generally do not restrict ourselves with the case 
$\gamma\geq0$, which corresponds to the presence of the viscous friction in
the continuum sub-system.

In the paper, the case of the system under consideration, where
\begin{equation}
\varGamma=0, \qquad \gamma=0
\label{nondiss-ass}
\end{equation}
is referred to as ``the non-dissipative case''. The case, when 
Eq.~\eqref{nondiss-ass} is not fulfilled,
i.e.,
\begin{equation}
  \varGamma\neq0\quad\text{or}\quad\gamma\neq0 
\label{diss-ass}
\end{equation}
is referred to as ``the dissipative case'', even if $\varGamma<0$ and/or
$\gamma<0$.

Though, in our previous papers 
\cite{Gavrilov2022jsv,Gavrilov2019nody,Gavrilov2019asm,indeitsev2016evolution,gavrilov2002etm},
it was assumed that we dealt with a transverse
oscillation of a taut string, now we consciously do not specify the
physical nature of the continuum sub-system. One of the
reasons is that the case when $\mathcal T$ depends on 
$\epsilon x$ 
does not have a simple interpretation in the framework of such a model (and does
have if we deal with a longitudinal or rotational oscillation). 
Accordingly, we
do not specify dimensions, which can be different depending on the
interpretation, when working with dimensional physical quantities. 
We do not use non-dimensional formulation since it is not useful in the context of
problems with several varying parameters. Another
reason is that in previous papers 
\cite{Gavrilov2022jsv,Gavrilov2019nody,Gavrilov2019asm,indeitsev2016evolution,gavrilov2002etm}
we used various kinds of the
non-dimensionalization.

%
%

At the instant $t=0$ an external given force $p(t)$ emerges, which is applied to the discrete
sub-system. Simultaneously, the discrete sub-system  starts to move along the
string according to the following law:
\begin{equation}
\label{l(t)}
{\l}(t) = {\l}(0) + \int_0^t {\v}(\epsilon\ttp)\,\d \ttp.
\end{equation}
Here,  $\l(0)$ is a given initial position for the discrete sub-system and 
\begin{equation}
v=v(\epsilon t)
\label{v(et)}
\end{equation}
is the speed of the sub-system (a given smooth slowly time-varying function). 
We assume  a sub-critical regime for the motion of the discrete sub-system, i.e.,
for all $t$ the following inequality must be satisfied in a certain
neighbourhood of $x=\l(t)$:
\begin{equation}
\label{v(t)and_c}
\abs{\v(\epsilon t)}<c(\epsilon x,\epsilon t),
\end{equation}
where
\begin{equation}
c=\sqrt{\dfrac{\TF}{\rho}}
\label{c-def}
\end{equation}
is the local instantaneous value for the speed of the wave propagation (the speed of
sound).
The discrete and continuum sub-systems are kinematically coupled, namely, it
is assumed that 
\begin{equation}
\label{U}
\UF(t) = u(\l(t),t).
\end{equation}
Accordingly, the unknown internal force on the continuum sub-system is
expressed by the term  $-P(t)\delta(x-\l(t))$ in the right-hand side of
Eq.~\eqref{eq1}, where
$\delta(\cdot)$ is the Dirac delta-function. Thus, for $v\neq 0$ our problem
transforms into a kind of moving load problem \cite{fryba1972vibration}.


Consider Eqs.~\eqref{eq2}, \eqref{eq1} in the co-moving with the discrete
sub-system co-ordinates:
\begin{equation}
\xi = x-\l(t),\qquad \tau=t. 
\label{co-moving}
\end{equation}
One has
\begin{equation}
\begin{gathered}
\pd {} x = (\cdot)', \qquad \pd {} t= \dot{(\cdot)} - v(\cdot)';
\\
\pdd {} x = (\cdot)'', \qquad \pdd {} t= \ddot{(\cdot)} + v^2(\cdot)''
-2v\dot{(\cdot)}'-\dot v(\cdot)',
\qquad
\ppdd{u}{x}{t}=\dot{(\cdot)}'-v(\cdot)'',
\end{gathered}
\label{to-co}
\end{equation}
where prime and overdot denote the partial derivatives with respect to $\xi$
and $\tau$, respectively.
Now, equations \eqref{eq2}, \eqref{eq1} can be rewritten as
\begin{gather}
\label{eq2m}
( M {\dot \UF})^\bcdot 
{+2\epsilon\GA \dot{\UF}}
+ K \UF = -P(\tau) +p(\tau),
\\
\label{eq1m}
(\TF -\rho\v{^2} )u''
+\big(
(\TF -\rho\v{^2})'  
+ 
({\rho}\v)^\bcdot
\big) 
u'
+ 2 \rho v \dot u '
- (
\dot \rho
- \rho' v
)
\dot u
- \rho \ddot u
{-2\epsilon \gamma(
\dot u -v u'
)}  - k u = -P(\tau)\delta(\xi),
\end{gather}
where
\begin{equation}
\label{U-tau}
\UF(\tau) = u(0,\tau).
\end{equation}

According to formulae, which are inverse of Eq.~\eqref{to-co}, 
 for an arbitrary quantity $\mu(\epsilon x, \epsilon t)$
one has
\begin{equation}
\mu'=O(\epsilon),\qquad \dot\mu=O(\epsilon),\qquad
\mu''=O(\epsilon^2),\qquad \ddot\mu=O(\epsilon^2), \qquad
\dot\mu'=O(\epsilon^2),
\end{equation}
provided that
Eq.~\eqref{v(et)} is true. Thus, provided that 
Eqs.~\eqref{all-slow},
\eqref{v(et)} are true, we can assume that quantities in Eq.~\eqref{eq1m} are
such that
\begin{equation}
\mathcal T=\mathcal T(\epsilon \xi, \epsilon \tau),
\qquad
\rho=\rho(\epsilon \xi, \epsilon \tau),
\qquad
k=k(\epsilon \xi, \epsilon \tau),
\qquad
\gamma=\gamma(\epsilon \xi, \epsilon \tau).
\label{mall-slow}
\end{equation}
In what follows in the paper, we deal with basic equations 
\eqref{eq2m}, \eqref{eq1m}
formulated in the co-moving co-ordinates.

The following Hugoniot conditions must be satisfied at $\xi=0$:
\begin{gather}
\label{Hugoniot1}
[u] = 0,\\
\label{Hugoniot2}
\left[u' \right] = -\frac{P(\tau)}{\TF - \rho \v{^2}}.
\end{gather}
Here and in what follows $[\mu]\equiv \mu(\xi+0)-\mu(\xi-0)$ 
for any arbitrary quantity $\mu(\xi,\tau)$.
The initial conditions for Eq. \eqref{eq1m} can be formulated in the following form, which
is conventional for distributions (or generalized functions) \cite{Vladimirov1971}:
\begin{equation}
\label{incon}
u\big|_{\tau<0}\equiv 0.
\end{equation}

In the paper, we assume that the external loading $p$ is a pulse, which
acts during some time, i.e., a given real integrable function 
(or a generalized
function \cite{Vladimirov1971}) of compact support
\cite{Nikolskii}:
\begin{equation}
p(\tau)\equiv 0, \qquad \tau<0 \quad\text{or}\quad \tau>\tau_0
\label{pulse}
\end{equation}
for certain $\tau_0>0$. 

\begin{remark} 
In \cite{Gavrilov2019nody} (see Sect.~\ref{Sect-Gavrilov2019nody}) 
we apply our method in the case, where 
$p(\tau)$ is an exponentially vanishing function and get an excellent
agreement. Thus, in principle, one can try to use our results for a wider class
of external pulse loadings.
\end{remark}

In zeroth order approximation $\epsilon=0$, Eq.~\eqref{eq1} transforms
into a
linear Klein-Gordon equation with constant coefficients. The latter equation 
in the co-moving co-ordinate system \eqref{co-moving}, 
namely Eq.~\eqref{eq1m}, has alternated coefficients and 
an additional term $2\rho v\dot u'$ in the left-hand side compared with
Eq.~\eqref{eq1}.  Equation \eqref{eq2m}
coincides with
Eq.~\eqref{eq2}, wherein $t=\tau$. 
Considering Eqs.~\eqref{eq2m}, \eqref{eq1m} at $\epsilon=0$, we see 
that the properties of the corresponding system can be completely
characterized by an $n$-tuple $\pp$
of six ($n=6$) real parameters,
namely
\begin{equation}
\begin{gathered}
\pp\=(\p_1,\p_2,\p_3,\p_4,\p_5,\p_6)\in\mathbb R^6;\\
\p_1=K,
\quad
\p_2=M,
\quad
\p_3=\TF,
\quad 
\p_4=\rho,
\quad 
\p_5=k,
\quad
\p_6=v.
\end{gathered}
\label{pP}
\end{equation}
Under certain conditions (the localization conditions, see Remark~\ref{loc-remark}), 
which are fulfilled in a certain domain (the localization domain) 
$\mathbb L\subset \mathbb R^6$ in the
six-dimensional problem parameter space, the free
oscillation of the mass-spring system remains non-vanishing, as we
observe in the case of an isolated discrete mass-spring system.
This happens to due
to the existence of a unique trapped mode, to which a spatially
localized near the discrete sub-system non-stationary oscillation corresponds, 
see Appendix~\ref{App-trapped}.
{For $\xi=0$, provided that the external loading is a pulse,  
this localized
oscillation asymptotically dominates over all other motions} 
\cite{Shishkina2023jsv}.
The frequency of this localized oscillation is $\Omega_0>0$ defined by
Eq.~\eqref{FR_0}.
The necessary auxiliary
results related to the stationary and non-stationary problems concerning a localized
oscillation in the system described by Eqs.~\eqref{eq2m}--\eqref{eq1m} are
derived, e.g.,
in \cite{Gavrilov2022jsv}. For convenience of the reader,
in {Appendix~\ref{app-dispersion}, we reproduce 
the basic formulae in the dimensional form used in the present paper.

To characterize completely all variable properties of the system described by 
Eqs.~\eqref{eq2m}, \eqref{eq1m} in the case $\epsilon>0$, additionally to $\pp$,
we need to introduce
a $\tilde n$-tuple $\ppp$ ($\tilde n=2$) 
\begin{equation}
\ppp\=(\tilde\p_1,\tilde\p_2)\in\mathbb R^2;\qquad
\tilde\p_1=\varGamma,\quad \tilde\p_2=\gamma.
\label{pPP}
\end{equation}
In the latter case, the tuples $\pp$ and $\tilde\pp$ are functions of
$\epsilon x$ and $\epsilon\tau$.
\begin{remark}	
\label{remark-small-diss}
We consider the case of a small dissipation only, since we want the existence
of the trapped mode to be possible in the zeroth order system with $\epsilon=0$.
\end{remark}

We plan to evaluate the displacements $\mathcal U(\tau)=u(0,\tau)$
of the discrete subsystem only. Namely, we look for the asymptotic solution
for $\mathcal U$ 
under the following conditions:
\begin{itemize}
\item 
$
\epsilon=o(1);
$
\item
$\tau=O(\epsilon^{-1}),$ thus, 
\begin{equation}
T\=\epsilon \tau=O(1);
\end{equation}
\item
The instantaneous values of the
zeroth order system parameters are such that 
\begin{equation}
\pp(\epsilon\xi,\epsilon \tau)
\in\LL\subset\mathbb R^6
\label{s-parameters}
\end{equation}
for $\xi$ from certain neighbourhood of zero
and all $\tau\geq0$.
\end{itemize}
Provided that these assumptions are true, we assume that the localized
oscillation with frequency $\Omega_0(T)$ is
an asymptotically dominant component of $u(0,\tau)$ as well as it is for
$\epsilon=0$.
The frequency $\Omega_0(T)$ is defined by the same equation \eqref{FR_0}, as well
as in the case $\epsilon=0$, wherein
the instantaneous values of system parameters, {taken at $\xi=0$ for the continuum
sub-system}, are under
consideration.

If the instantaneous value of the vector of the zeroth order system
parameters \eqref{s-parameters}
at $\xi=0$ leaves the localization domain, then we expect very fast
vanishing of the displacements $\mathcal U(t)$. In a particular case, this was
demonstrated numerically in \cite{Gavrilov2022jsv}. Later, we have shown that
at least for $\epsilon=0$
the displacement $\mathcal U(t)$ in the composite system vanishes
asymptotically faster than
in the corresponding pure continuum system. This happens due to the
existence of the wave phenomenon, which we have called the anti-localization of
non-stationary linear waves \cite{Gavrilov2024,Gavrilov2023DD,Shishkina2023cmat,Shishkina2023jsv}.

\begin{remark}  
In \cite{Gavrilov2022jsv} (see Sect.~\ref{Sect-Gavrilov2022jsv}), besides a free localized oscillation, it was
considered the
forced oscillation caused by a class of loadings, which 
can be represented as a superposition of a pulse $\hat p(\tau)$, which
satisfies 
Eq.~\eqref{pulse},
and a number of harmonics with slowly time-varying
frequencies and  amplitudes:
\begin{equation}
  p(\tau,T)=H(\tau)\left(\frac{\hat p(\tau)}2+\sum_{i=1}^Np^{(\Omega_i)}(T)\exp\left(-\I\int_0^\tau
  \Omega_i(\tp)\,\d \tp
  \right)\right)+\mathrm{c.c.},
\label{p-def}
\end{equation}
where the notation {$\mathrm{c.c.}$ denotes 
the complex conjugate terms for the whole right-hand side, $H(\cdot)$ is the
Heaviside step-function},
\begin{gather}
\Omega_i\geq0,
\\
\Omega_i\not\simeq\Omega_j \quad \forall t\quad
\text{if} \quad{i\neq j}\quad\text{for}\quad{i,j=\overline{1,N}}.
\label{non-resonant}
\end{gather}
Here, the amplitudes $p^{(\Omega_i)}(T)\ (i=\overline{1,N})$ are given
smooth complex-valued functions.
The results concerning the forced oscillation can be straightforwardly
transferred to the problem under consideration in this paper. On the other
hand, taking into account the forced oscillation 
complicates the calculations. Thus, in this paper, we deal with pulse loadings
and a free localized oscillation only. 
\end{remark}

\begin{remark}	
The special limiting case 
\begin{equation}
\rho=0
\label{rho0}
\end{equation}
corresponds to a discrete sub-system attached to an inertialess waveguide. In
this case, the continuum sub-system behaves as an additional distributed
spring attached to the discrete sub-system. The similar models were used in
studies \cite{smith1964motions,dyniewicz2009paradox,Gavrilov2016nody}.
\end{remark}
\begin{remark}	
The formal 
limiting case when 
\begin{equation}
\mathcal T=0,\qquad \rho=0, \qquad k=0, \qquad \gamma=0
\label{case-no-co}
\end{equation}
corresponds to an isolated uncoupled mass-spring-damper system with
time-varying properties described by Eq.~\eqref{eq2} with $P=0$. A free oscillation in such a system is considered in 
Appendix~\ref{App-1d}. We recommend the reader to look through the material
in Appendix~\ref{App-1d} before reading the rest of the paper.
\end{remark}

\section{Derivation of the first approximation equation}
\label{sect-approach}

 

For $\xi>0$ and $\xi<0$ we look for the solution in the form, which we called 
in \cite{Gavrilov2022jsv} ``the
single-frequency ansatz'', or better to say, ``the single mode ansatz'' corresponding to the trapped mode frequency $\Omega_0(T)$:
\begin{gather}
u(\xi,\tau)= W(X,T)\exp
\phi(\xi,\tau)
+\cc,
\label{ansatz-u}
\end{gather}
where 
\begin{gather}
X=\epsilon \xi,
\qquad 
T=\epsilon \tau
\end{gather}
are the slow spatial co-ordinate and the slow time; $\phi(\xi,\tau)$ such that
\begin{gather}
\phi' = \I\omega(X,T), \qquad \dot{\phi}=-\I\Omega(X,T),
\label{OSC-fast-phases}
\\
\lim_{X\to0}\Omega=\Omega_0(T)
\label{lim-Omega}
\end{gather}
is the fast phase;
\begin{gather}
W(X,T)=\sum\limits_{j=0}^\infty \epsilon^j W_j(X,T)
\label{expansion-basic}
\end{gather}
such that
\begin{gather}
\mathcal W(T)\=\lim_{X\to0} W(X,T),
\\
\mathcal W_j(T)\=\lim_{X\to0} W_j(X,T)
\label{lim-W}
\end{gather}
is the amplitude. 
The wave-number $\omega(X,T)$ and the frequency $\Omega(X,T)$
should satisfy dispersion relation 
\eqref{dis_rel}
 and equation 
\begin{equation}
{\Omega}'_X+{\omega}'_T=0
\label{OSC-dxx-SPRING}
\end{equation}
that follows from
\eqref{OSC-fast-phases}
for all $X$ and $T$ in a neighbourhood of $X=0$.
In this case, the phase $\phi(\xi,\tau)$ can be defined by the formula 
\begin{equation}
 \phi=\I\int(\omega\,\d\xi-\Omega\,\d\tau).
\end{equation}
Additionally, we require that 
\begin{gather}
[W_j]=0,\qquad 
[\phi]=0.
\label{jumps-Wphi}
\end{gather}

\begin{remark}	
Equation~\eqref{OSC-dxx-SPRING} can be recognized as the space-time eikonal
equation, see Eq.~(A.6.3) in
\cite{Babich2009}.
\end{remark}

Note that the analytic expressions for
quantities 
$u,\ W,\ W_j,\ \phi,\ \varOmega,\ \omega$
are generally different for $\xi\lessgtr0$. Therefore, we
additionally require that $u$ satisfy vanishing boundary conditions at infinity
($\xi\to\pm\infty$). Accordingly, we need to choose for $\xi\gtrless0$
different roots
\eqref{omega}
of dispersion relation
\eqref{dis_rel}:
\begin{gather}
u(\xi,\tau)=\sum_{(\pm)} H(\pm\xi)\,W^{(\pm)}(X,T)\exp
\phi^{(\pm)}(\xi,\tau)+\cc.
\label{ansatz-u-ext}
\end{gather}
In Eqs.~\eqref{ansatz-u}--\eqref{ansatz-u-ext} 
and in what follows, we drop superscript $(\pm)$ for the aim of simplicity.
For $\epsilon=0$ the single-mode ansatz 
{\eqref{ansatz-u}--\eqref{ansatz-u-ext}} should transform into the solution
\eqref{eq:solution_inhomogeneous_stationary_problem-ext}
of the corresponding zeroth order problem with an accuracy to an unknown
constant multiplier. 

\begin{remark} 
The structure of the single-mode ansatz \eqref{ansatz-u}--\eqref{ansatz-u-ext}
assumes that
\begin{itemize}
\item Dispersion relation~\eqref{dis_rel} holds for all $T$ and
$X$, in particular for $X=\pm0$;
	
\item Frequency equation~\eqref{fr_eq} for the trapped mode holds for all $T$.
\end{itemize}
\end{remark}

According to the procedure of the method of multiple scales \cite{nayfeh2008perturbation},
the slow variables $X$, $T$, and the fast phase $\phi$ are assumed to be independent
variables. In this way, we represent the differential operators with respect to time and the spatial co-ordinate in the following form:
\begin{equation}
\label{dot-xt}
\begin{gathered}
\dot{(\cdot)}=-\I \varOmega \partial_\phi + \epsilon \partial_T, \qquad 
(\cdot)'=\I\omega \partial_\phi +\epsilon \partial_X, \\
\ddot{(\cdot)}=-\varOmega^2 \partial_{\phi\phi}^2 - 2\epsilon \I \varOmega \partial_{\phi T}^2
   -\epsilon \I \varOmega'_T \partial_\phi +O(\epsilon^2), \\
(\cdot)'' = -\omega^2 \partial_{\phi\phi}^2 +2\epsilon \I \omega \partial_{\phi X}^2 
   +\epsilon \I \omega'_X \partial_\phi +O(\epsilon^2),\\
(\cdot)\dot{{\;}}'= \omega \varOmega \partial^2_{\phi \phi}-\epsilon \I
   \varOmega \partial_{\phi X}^2- \epsilon \I \varOmega'_X \partial_\phi
   +\epsilon \I \omega \partial^2_{\phi T}+O(\epsilon^2).
\end{gathered}
\end{equation}

To derive the first approximation equation, we substitute
the single-mode ansatz \eqref{ansatz-u}--\eqref{ansatz-u-ext}
as well as the representations for differential operators
\eqref{dot-xt} into three equations. These are 
\begin{description}       
\item[1)]
The second Hugoniot condition
\eqref{Hugoniot2}, where quantity $P(\tau)$ in the right-hand side is expressed
according to Eq.~\eqref{eq2m}; 
and
\item[2,3)]
{PDE} under investigation in the form of 
Eq.~\eqref{eq1m} considered at $\xi\to-0$ and $\xi\to+0$ (two additional
equations). 
\end{description}
For all resulting equations, 
the corresponding zeroth order approximation is satisfied
automatically. The joint consideration of these three equations allows one to
derive the first approximation equation in the form of an ODE.

At first, consider the second Hugoniot condition
\eqref{Hugoniot2} for $\tau>\tau_0$, where $\tau_0$ is defined by Eq.~\eqref{pulse}.
The internal force $P(\tau)$
in the right-hand side can be expressed
according to Eq.~\eqref{eq2m} wherein $p=0$.  Thus, we get
\begin{equation}
\label{eq:gugonio_in_non_stat}
[\I\omega W  + \epsilon W'_\X ]  + O (\epsilon^2)  
=\left.\frac{ M\left(-\O_{0}^2 W  -2 \epsilon \I\O_{0} {W}'_{\T} -
	\epsilon \I\O_{0}{}'_T W\right) -\epsilon \I M'_\T\O_0 W
	-{  2\epsilon\I \varGamma \Omega_0 W}+ K W}{\TF-\rho\v^2}
\right|_{X=0}.
\end{equation}
After equating of the  coefficients of like powers $\epsilon$ one can find
that the zeroth order approximation is
identically satisfied due to the
frequency equation \eqref{fr_eq}. The first approximation is
\begin{equation}
\label{W_0}
\left.
[ W_{0}{}'_{\X} ] = 
-\frac{2\I M \Omega_{0} {W_0}'_T +  
\I \big(M{\O_0{}'_\T +   M'_T\O_{0} { + 2\varGamma \Omega_0}\big) W_{0} }}
{\TF-\rho\v^2}
\right|_{X=0}.
\end{equation}
Note that the last equation does not involve the terms, which depend on $W_{1}$,
since the procedure of the method of multiple scales guaranties that the
common multiplier before all such terms equals zero due to the
frequency equation \eqref{fr_eq}. The left-hand side of Eq.~\eqref{W_0} equals
the magnitude of the jump discontinuity $[W_0{}'_X]$, whereas the right-hand side
does not involve any spatial derivatives.

Now we plan to calculate
$[W_0{}'_X]$ in an independent way using PDE 
\eqref{eq1m} considered at $\xi=\pm0$ (or, equivalently, at $X=\pm0$).
To do this, we substitute \eqref{ansatz-u} and \eqref{dot-xt} into 
Eq.~\eqref{eq1m}, taking into account that the right-hand side of
Eq.~\eqref{eq1m} is zero for all $\xi\neq0$.
After equating of the  coefficients of like powers $\epsilon$, one can find
that zeroth order approximation is
identically satisfied due to the
dispersion relation 
\eqref{dis_rel}.
For the first order
approximation, we obtain:
\begin{multline}
\big(( \TF - \rho \v^2) 2\omega - 2\rho\v \O_0 \big) W_0{}'_{\X} + 
\rho ( 2\v\omega + 2 \O_{0}) W_0{}'_{\T}  \\
+
 \left( (  \TF - \rho \v^2 ) \omega'_\X +  2\rho\v \omega'_\T +
 \rho\O_0{}'_{\T} + 
 \big((\TF - \rho\v^2 )'_{\X} + ( \rho\v)'_{\T}
 \big)\omega
+ ( \rho'_{\T} - \rho'_{\X} \v ) \Omega_0  
+2\gamma (\Omega_0+v\omega)\right) W_0
= 0
\label{first-app-1}
\end{multline}
or
\begin{multline}
\big(( \TF - \rho \v^2 ) 2\omega - 2\rho\v \O_0\big) W_0{}'_\X +
\left(2\rho\v \omega + 2\rho \O_{0} \right) W_0{}'_{\T} 
\\
+
\left( \big(( \TF - \rho \v^2 ) \omega -  \rho\v\O_0 \big)'_\X
+ ( \rho\v\omega + \rho\Omega_0 )'_{\T} 
+2\gamma (\Omega_0+v\omega) \right) W_0 = 0.
\label{first-app-2}
\end{multline}
Again, the last equation does not involve the terms, which depend on $W_{1}$,
since the procedure of the method of multiple scales guaranties that
the common multiplier before all such terms equals zero due to the
dispersion equation 
\eqref{dis_rel}.
Using Eqs.~\eqref{omega}, \eqref{B} 
we obtain:
\begin{multline}
\left( \TF - \rho \v^2 \right) \omega -  \rho\v\O =
\left( \TF - \rho \v^2 \right)B\pm \I\left( \TF - \rho \v^2 \right) S-  \rho\v\O
=
\left( \TF - \rho \v^2 \right)\frac{\rho\v\O}{\TF-\rho\v^2}
\pm \I\left( \TF - \rho \v^2 \right) S-  \rho\v\O
\\=
\rho\v\O
\pm \I\left( \TF - \rho \v^2 \right) S-  \rho\v\O 
=
\pm \I\left( \TF - \rho \v^2 \right) S.
\end{multline}
Thus, we can rewrite Eq.~\eqref{first-app-2} as follows:
\begin{multline}
\label{first-app-3}
\pm 2\I( \TF - \rho \v^2)S \, W_0{}'_{\X} +
2 \left( \rho\v \omega + \rho \O_{0} \right) W_0{}'_{\T} 
\\
+
\left( \left(\pm \I( \TF - \rho \v^2 ) S \right)'_\X
+
\left( \rho\v\omega + \rho\Omega_0 \right)'_{\T} 
+2\gamma (\Omega_0+v\omega) \right) W_0
= 0
\end{multline}
or
\begin{equation}
{W_0}'_{\X}
= - \frac{ \left( 2\rho\v \omega + 2\rho \O_{0} \right) {W_0}{}'_{\T} + 
	\left( \big(\pm \I( \TF - \rho \v^2 ) S \big)'_\X
	+
	\left( \rho\v\omega + \rho\Omega_0 \right)'_{\T} 
+2\gamma (\Omega_0+v\omega) \right) W_0}
{	\pm 2\I( \TF - \rho \v^2 ) S }.
 \label{eq:equation}
\end{equation} 
Taking into account the eikonal equation \eqref{OSC-dxx-SPRING} and Eq.~\eqref{omega}, we get
\begin{multline}
\pm \I\big(( \TF - \rho \v^2 ) S \big)'_\X =
- \left( \TF - \rho \v^2 \right) 
(\pm  \I S'_{\O} B'_{\T} - S'_{\O}S'_{\T} ) 
\pm \I\big(( \TF - \rho \v^2 ) S \big)'_{\rho} \rho'_{\X}  
\\
{ \pm \I\big(( \TF - \rho \v^2 ) S \big)'_{\TF} \TF'_{\X}}
{ \pm \I\big(( \TF - \rho \v^2 ) S \big)'_k k'_{\X}}.
\label{new-identity}
\end{multline}

\begin{remark}  
We remind that formulae 
\eqref{first-app-1}--\eqref{new-identity} are written for $X=\pm0$.
\end{remark}
Calculating the magnitude
$[ W_0{}'_\X ]$
of jump discontinuity yields:
\begin{equation}
\left.
[ W_0{}'_\X ]= - \frac{\left(2\rho\v  B	+ 2\rho \O_{0}\right){W_0'}_\T + \left(	\left( \TF - \rho \v^2 \right) S'_{\O}S'_{\T} + 
\left( \rho\v B + \rho\O_{0} \right)'_{\T} {
+2\gamma (\Omega_0+vB)}\right) W_{0}}{	\I\left( \TF - \rho \v^2 \right) S}
\right|_{X=0}.
\label{jump-dX}
\end{equation} 

Finally, equating the right-hand sides of Eqs.~\eqref{W_0} and \eqref{jump-dX}
results in the first approximation equation for $\WF_0(T)$ defined by 
Eq.~\eqref{lim-W}
in the
form, which does not involve any spatial derivatives:
\begin{multline}
\frac{2 M \Omega_{0} {\WF_0}'_T +   (M{\O_{0}}'_{\T} +   M'_T\O_{0}
{+ 2\varGamma \Omega_0}) \WF_{0} }{(\TF-\rho\v^2)}
=\\
- \frac{\left(2\rho\v  B	+ 2\rho \O_{0}\right){\WF_0}'_\T + \left(
  \left( \TF - \rho \v^2 \right) S'_{\O}S'_{\T} + 	\left( \rho\v B +
  \rho\O_{0} \right)'_{\T} {+2\gamma(v B+\Omega_0)}\right) \WF_{0}}{	\left( \TF - \rho \v^2 \right) S}.
\label{before-solving}
\end{multline}

One has $\TF\neq\rho\v^2$,  
due to Eqs.~\eqref{v(t)and_c}, \eqref{c-def},
and $S\neq0$ due to 
Eqs.~\eqref{S}, \eqref{interval}. Thus,
Eq.~\eqref{before-solving} can be 
rewritten in the form
\begin{multline}
{2 M \Omega_{0}S {\WF_0}{}'_T +   (M\O_{0}{}'_{\T} +   M'_T\O_{0}{+ 2\varGamma \Omega_0})\,S \,\WF_{0} }
=\\
- {\left(2\rho\v  B	+ 2\rho \O_{0}\right){\WF_0}'_T - \left(	\left( \TF - \rho \v^2 \right) S'_{\O}S'_{\T} + 	\left( \rho\v B + \rho\O_{0} \right)'_{\T}
{+2\gamma(v B+\Omega_0)}\right) \WF_{0}},
\end{multline}
which is equivalent to
\begin{multline}
{\left(2\rho\v  B	+ 2\rho \O_{0} + 2 M \Omega_{0}S \right) \WF_0{}'_T}
=\\
- {\left(	\left( \TF - \rho \v^2 \right) S'_{\O}S'_{\T} + 	\left(
  \rho\v B + \rho\O_{0} \right)'_{\T} + (M{\O_{0}'}_{\T} +
  M'_T\O_{0})\,S
  {+ 2\varGamma \Omega_0\,S+2\gamma(v B+\Omega_0)}\right) \WF_{0}}
\end{multline}
or
\begin{multline}
{\left(2\rho\v  B	+ 2\rho \O_{0} + 2 M \Omega_{0}S \right) {\WF_0}'_T}
=\\
- {\left(	\left( \TF - \rho \v^2 \right) S'_{\O}S'_{\T} 
- M\O_{0}S'_T
+ 
\left( \rho\v B + \rho\O_{0}  + M\O_{0} S \right)'_{\T}  
{+ 2\varGamma \Omega_0\,S+2\gamma(v B+\Omega_0)}\right) \WF_{0}}.
\end{multline}
The last equation can be transformed to the following one:
\begin{equation}
\label{WW}
\frac{ {\WF_0}'_T}{\WF_{0}} =
-\frac{1}{2} \frac{\left(\TF-\rho\v^2 \right)S'_{\O}-M\O_{0}}
 {\rho\v B +\rho \O_{0} + M\O_{0}  S } S'_{\T} 
-\frac{1}{2} \frac{\left( \rho\v B + \rho\O_{0} +M\O_{0} S \right)'_{\T} }
{\rho\v B+\rho \O_{0} +M\O_{0} S }
-{ \frac{\varGamma \Omega_0\,S+\gamma(v B+\Omega_0)}
{\rho\v B + 	\rho \O_{0} +M\O_{0} S}}.
\end{equation}

\section{Analysis of possible approaches to solve the first approximation equation}
\label{Sect-general}

By construction, the first approximation equation
\eqref{WW}, as well as Eq.~\eqref{App1-1st-P} considered in
Appendix~\ref{App-1d}, can be equivalently transformed to the equation of the
following structure:
\begin{gather}
\frac{\WF_0{}'_T}{\WF_0}
=
\label{gen-K-M-1st}
\sum_{i=1}^n F_i(\pp)\,\p_i{}'_T
+\left\{
\sum_{i=1}^{\tilde n}\tilde F_i(\pp)\,\tilde\p_i{}
\right\},
\end{gather}
Here $F_i(\pp),\ \tilde F_i(\pp)$ are certain functions of 
$\pp$; $\pp$ and $\tilde\pp$ are system parameters introduced by 
Eqs.~\eqref{pP}, \eqref{pPP}.
The first sum in the right hand-side of Eq.~\eqref{gen-K-M-1st} emerges 
since the coefficients of Eqs.~\eqref{eq2m}, \eqref{eq1m}
are slowly time-varying
functions. The second sum in the curly brackets is due to coefficients 
of Eqs.~\eqref{eq2m}, \eqref{eq1m}, which are 
proportional to~$\epsilon$. Provided that the following conditions
\begin{gather}
\left\{
\sum_{i=0}^{\tilde n}\tilde F_i(\pp)\,\tilde\p_i{}
\right\}'_T
=0
\quad\Leftrightarrow\quad
\sum_{i=0}^{\tilde n}\tilde F_i(\pp)\,\tilde\p_i{}
=\Lambda=\text{const},
\label{PD1}
\\
\pd{F_i}{\p_j}=\pd{F_j}{\p_i},\quad i\neq j,
\label{PD2}
\end{gather}
are satisfied, 
Eq.~\eqref{gen-K-M-1st} 
can be rewritten in the form of the equation
\begin{equation}
\frac{\WF_0{}'_T}{\WF_0}
=
{J'_T\big(\pp)}+
\Lambda
,
\label{exact-d-lambda}
\end{equation}
where the expression in the right-hand side is the exact derivative of a
function
$
J(\pp)+
\Lambda T
.
$
Now,
Eq.~\eqref{exact-d-lambda}
can be integrated:
\begin{equation}
\WF_0=\mathcal C\,\Exp{J(\pp)+
\Lambda T
}.
\end{equation}
Here $\mathcal C$ is an arbitrary complex constant.
Thus, the solution is obtained in the form of a function, depending on
the instantaneous values of the system parameters $\pp(0,T)$ and the current slow time
$T$ in the explicit way.
Such a solution is valid for
an arbitrary history of $\pp(0,\tp)$, $\tp\leq T$. In the particular case
\begin{equation}
\Lambda=0,
\label{Lambda-zero}
\end{equation}
Eq.~\eqref{gen-K-M-1st} 
can be rewritten as
\begin{equation}
\frac{\WF_0{}'_T}{\WF_0}
=
J'_T\big(\pp),
\label{exact-d-P}
\end{equation}
and the corresponding solution can be obtained in the form of a function, depending on
the current values of the system parameters $\pp(0,T)$ only:
\begin{equation}
\WF_0=\mathcal C\,\exp{J(\pp)}.
\label{q11}
\end{equation}
If $\Lambda$ is not a constant, i.e.,
condition 
\eqref{PD1} is not satisfied,
the solution of 
Eq.~\eqref{exact-d-lambda} is
the functional
\begin{equation}
\WF_0=\mathcal C\,\Exp{J(\pp)+
\int_0^T
\sum_{i=1}^{\tilde n}\tilde F_i\big(\pp\big)\,\tilde\p_i{}
\,\d \tp
},
\label{q1}
\end{equation}
which depends on the history of $\pp(0,\tp)$ and $\tilde\pp(0,\tp)$. Finally, if both conditions 
\eqref{PD1} and
\eqref{PD2}
are not satisfied,  the solution of 
Eq.~\eqref{gen-K-M-1st} is
the functional
\begin{equation}
\WF_0=\mathcal C\,\Exp{\int_0^T 
\left(
\sum_{i=1}^n F_i(\pp)\,\p_i{}'_T
+
\sum_{i=1}^{\tilde n}\tilde F_i(\pp)\,\tilde\p_i{}
\right)
\,\d \tp
},
\label{q2}
\end{equation}
which again depends on the history of $\pp(0,\tp)$ and $\tilde\pp(0,\tp)$.
In last two cases
\eqref{q1} and \eqref{q2}, if the history is not
specified, the
solution of \eqref{gen-K-M-1st} can be found 
in quadratures only.
If the history of $\pp(0,\hat T),\ \tilde\pp(0,\hat T)$ is
known, the solution is an explicit function of slow time~$T$.

Now, let us note that for reasons of convenience, it may be
useful to use in calculations, instead of $\pp$,
some new variables: 
\begin{equation}
\boldsymbol\Psi\=\big(\Psi_1(\pp),\dots,\Psi_\nu(\pp)\big),
\label{bPsi-def}
\end{equation}
where $\nu$ is not necessary equal to $n$. 
The first approximation for the amplitude 
\eqref{gen-K-M-1st}
in terms of variables $\bPsi$ is
\begin{gather}
\frac{\WF_0{}'_T}{\WF_0}
=
\label{gen-K-M-1st-psi}
\sum_{i=1}^\nu \Phi_i(\bPsi)\,\Psi_i{}'_T
+
\sum_{i=1}^{\tilde n}\tilde \Phi_i(\bPsi)\,\tilde\p_i{}
,
\end{gather}
where $\Phi_i(\bPsi),\ \tilde \Phi_i(\bPsi)$ are certain functions of
the variables $\bPsi$.
If Eq.~\eqref{gen-K-M-1st-psi} can be rewritten in the form
\begin{equation}
\frac{\WF_0{}'_T}{\WF_0}
=
{\J'_T\big(\bPsi)}+
\Lambda
,
\label{exact-d-psi}
\end{equation}
where
\begin{equation}
\Lambda=
\sum_{i=0}^{\tilde n}\tilde \Phi_i(\bPsi)\,\tilde\p_i{}
=\text{const},
\label{Lambda-is-constant}
\end{equation}
then conditions 
\eqref{PD2}
are definitely satisfied, and 
the solution can be again obtained in the form of a function, depending on
the current values of the system parameters $\pp(0,T)$ and current slow time
$T$ in the explicit way:
\begin{equation}
\WF_0=\mathcal C\,\Exp{\J\big(\bPsi(\pp)\big)+
\Lambda T
}
.
\label{sol-Lambda-is-constant}
\end{equation}
In the particular case
\eqref{Lambda-zero},
Eq.~\eqref{gen-K-M-1st} 
can be rewritten as
\begin{equation}
\frac{\WF_0{}'_T}{\WF_0}
=
\J'_T\big(\bPsi),
\label{exact-d}
\end{equation}
and the corresponding solution can be obtained in the form of a function, depending on
the current values of the system parameters $\pp(0,T)$ only:
\begin{equation}
\WF_0=\mathcal C\,\exp{\J(\bPsi)}.
\label{sol-d}
\end{equation}
Let us note that the possibility to represent a given equation with the
structure of Eq.~\eqref{gen-K-M-1st} in the form of 
Eq.~\eqref{exact-d-psi} or 
Eq.~\eqref{exact-d} is absolutely unobvious. Of course, one can try to check
conditions 
\eqref{PD2} directly. However, this can be a difficult problem due to a lengthy structure of
the right-hand side of the equation under consideration.
On the other hand, a lucky choice of the variables $\bPsi(\pp)$ can essentially 
simplify the
calculations and help to obtain the representation of Eq.~\eqref{gen-K-M-1st} in the
form of Eq.~\eqref{exact-d-psi} or Eq.~\eqref{exact-d}.

If $\Lambda$ is not a constant, i.e.,
condition 
\eqref{Lambda-is-constant}
is not satisfied,
the solution of 
Eq.~\eqref{exact-d-lambda} is
the functional:
\begin{equation}
\WF_0=\mathcal C\,\Exp{\J(\bPsi)+
\int_0^T
\sum_{i=1}^{\tilde n}\tilde \Phi_i\big(\bPsi\big)\,\tilde\p_i{}
\,\d \tp
}.
\label{sol-general}
\end{equation}

\begin{remark}	
\label{remark-one-parametric}
If the tuple $\tilde\pp$ is empty, and only one $\p_j$ is a function of time $T$,
whereas all other $\p_i$, $i\neq j$, are constants,
then Eq.~\eqref{gen-K-M-1st} can be rewritten as 
\begin{gather}
\frac{\WF_0{}'_T}{\WF_0}
=
F_j(\pp)\,\p_j{}'_T
\label{1P-eq}
\end{gather}
Accordingly, conditions \eqref{PD1}, \eqref{PD2} are satisfied automatically, and the
solution can be obtained in the form of the following function:
\begin{equation}
\WF_0=\mathcal C\,\Exp{
\int_0^{\p_j}
F_j(\p_1,\dots \hat\p_j,\dots\p_n)
\,\d\hat\p_j
}.
\label{1P-sol-t}
\end{equation}
Though Eq.~\eqref{1P-sol-t} always provides the solution, the corresponding
calculations can be
difficult. In practical applications, the calculations of the function
$F_j(\pp)$ and the expression in the right-hand side of 
\eqref{1P-eq} may be quite lengthy. Instead,
it may be useful to introduce
$n$-tuple of new variables $\bPsi$ such that Eq.~\eqref{gen-K-M-1st-psi} can be
rewritten in the form of the following equation
\begin{gather}
\frac{\WF_0{}'_T}{\WF_0}
=
\sum_{i=1}^\nu \Phi_i(\Psi_i)\,\Psi_i{}'_T,
\label{gen-K-M-1st-psi-1}
\end{gather}
where variables $\bPsi$ in the right-hand side are separated. The last
equation can be integrated as follows:
\begin{equation}
\WF_0=\mathcal C\,\Exp{
\sum_{i=1}^\nu 
\int_0^{\Psi_i}
\Phi_i(\hat\Psi_i)
\,\d\hat\Psi_i
}.
\label{1P-sol}
\end{equation}
\end{remark}

\section{Solving of the first approximation equation}
\label{Sect-solving}

The first approximation equation \eqref{WW}
has the structure of Eq.~\eqref{gen-K-M-1st-psi} wherein
$\bPsi$ are the variables defined by Eq.~\eqref{bPsi-def} with $\nu=2$:
\begin{gather}
\Psi_1=S,\qquad \Psi_2=\rho\v B + \rho\O_{0} +M\O_{0} S;
\label{Psi-hack}
\\
\Lambda=
-{ \frac{\varGamma \Omega_0\,S+\gamma(v B+\Omega_0)}
{\rho\v B + 	\rho \O_{0} +M\O_{0} S}}.
\end{gather}
%
Noticing that the following identity 
\begin{equation}
 \frac{	\left( \TF - \rho \v^2 \right) S'_{\O}-	M\O_{0}}
{\rho\v B +\rho \O_{0} + M\O_{0} S } = -\frac{1}{S},
\label{relationship}
\end{equation}
is true \cite{Gavrilov2022jsv}, see Appendix~\ref{App-i},
we, finally, rewrite the first approximation equation in the form of 
Eq.~\eqref{exact-d-psi}:
\begin{equation}
\label{WW_1}
\frac{ {\WF_0}'_T}{\WF_{0}} = \frac{1}{2} \frac{ S'_{\T}}{S} 
-\frac{1}{2} 
\frac{\left( \rho\v B + \rho\O_{0} +M\O_{0} S \right)'_{\T} }
{\rho\v B+\rho \O_{0}+M\O_{0} S }-
\left\{ \frac{\varGamma \Omega_0\,S+\gamma(v B+\Omega_0)}
{\rho\v B + 	\rho \O_{0} +M\O_{0} S}\right\}
\end{equation}
wherein
\begin{equation}
\mathcal J
=
\ln\sqrt{\frac{S}{\rho\v B + \rho \O_{0} + M\O_{0} S}}
.
\end{equation}

In the general (dissipative) case, the solution of Eq.~\eqref{WW_1} has the structure
of the functional defined by Eq.~\eqref{sol-general}:
\begin{equation}
	\WF_{0}=\Co\mathcal A(T)\exp\big(-\mathcal D(T)\big),
\label{final-evolution}
\end{equation}
where $\Co$ is an arbitrary complex constant, and the functions 
$\mathcal A(T)$ and $\mathcal D(T)$ are 
\begin{equation}
\mathcal A=\sqrt{\frac{S}{\rho\v B + \rho \O_{0} + M\O_{0} S}}
,
\qquad
\mathcal D=\int_0^T  
\frac{\varGamma \Omega_0\,S+\gamma(v B+\Omega_0)}
{\rho\v B + \rho \O_{0} + M\O_{0} S}\, \d \tp.
\label{WW_2}
\end{equation}
 Substituting \eqref{B} and
\eqref{S} into the expression \eqref{WW_2}, one gets the equivalent forms
\begin{equation}
\mathcal A=
	\sqrt{
		\frac{
			\sqrt{
				k\TF-k\rho\v^2-\TF\rho\O_{0}^2
			}
		}{
			\O_{0} M
			\sqrt{
				k\TF-k\rho\v^2-\TF\rho\O_{0}^2
			}
			+
			\TF \rho\O_{0}
		}
	},
\qquad	
	\mathcal D=\int_0^T  \frac{\varGamma \sqrt{
				k\TF-k\rho\v^2-\TF\rho\O_{0}^2
			}
+\gamma\TF}{ M
			\sqrt{
				k\TF-k\rho\v^2-\TF\rho\O_{0}^2
			}
			+
			\TF \rho
}\, \d \tp
	\label{eq:non_stationar_amplitude}
\end{equation}
or
\begin{equation}
\mathcal A=	
	\sqrt{
		\frac{
			\sqrt{
				\Omega^2_\ast-\O_{0}^2
			}
		}{
			\O_{0} M
			\sqrt{
				\Omega^2_\ast-\O_{0}^2
			}
			+
		\sqrt{\TF \rho}\O_{0}
		}
	},
\qquad
\mathcal D=
	\int_0^T  \frac{\varGamma \sqrt{
				\Omega^2_\ast-\O_{0}^2
			}
+\gamma\sqrt{\TF/\rho}}{M
			\sqrt{
				\Omega^2_\ast-\O_{0}^2
			}
			+
			\sqrt{\TF \rho}
}\, \d \tp,
	\label{eq:non_stationar_amplitude-ots}
\end{equation}
where $\Omega_\ast$ is the cut-off frequency given by Eq.~\eqref{cutoff}.
Taking into account the frequency equation \eqref{fr_eq}, we obtain one more
equivalent form
\begin{equation}
\label{WW_final}
\mathcal A=\sqrt{\frac{M\O_0^2 - K}{\O_0(M^2\O_0^2-K M +2\TF
\rho)}}
,\qquad
\mathcal D=
\int_0^T  \frac{\varGamma (M\Omega_0^2-K)+2\gamma
\TF}{M^2\Omega_0^2-KM + 2 \TF\rho}\, \d \tp.
\end{equation}

For a non-dissipative system where 
Eq.~\eqref{nondiss-ass} is fulfilled,
we obtain the solution for the amplitude in the form of an algebraic
expression \eqref{sol-d}, depending on
current values of the system parameters $\bPsi(\pp)$ only, and
not involving the slow time in an explicit way. Such a solution is valid for
an arbitrary history of $\pp(0,\hat T)$. For a dissipative system where
Eq.~\eqref{diss-ass} is true, 
and additionally
\begin{equation}
\dot \Lambda\not\equiv 0,
\end{equation}
the solution for the amplitude generally is a functional,
which depends on the history of $\pp(0,T)$ and $\tilde\pp(0,T)$. If the history is not
specified, the solution of Eq.~\eqref{WW_1}  can be found 
in quadratures only. 
If the history of $\pp(0,T)$ and $\tilde\pp(0,T)$ is
known, the solution is an explicit function of slow time~$T$. The analogous
results are valid for a mass-spring-damper system with time-varying
parameters (see Appendix~\ref{App-1d}).

\begin{remark}  
\label{remark-log}
The possibility to obtain the solution of Eq.~\eqref{WW} in the
non-dissipative case 
\eqref{nondiss-ass}
in the form of an algebraic expression follows not only from
the fact that the right-hand side of 
Eq.~\eqref{WW_1} 
is the exact derivative of some function, but, additionally, from the fact
that 
the function is the logarithm of an algebraic expression.
\end{remark}

One can see from Eq.~\eqref{WW_final} that in the
formal limiting case 
\eqref{case-no-co}
of an isolated mass-spring-damper system, we have
\begin{equation}
  \WF_{0} \to \frac \Co{\sqrt{M\Omega_0}}\Exp{-\int_0^T \frac \varGamma {M} \,\d \tp}.
\end{equation}
Since, due to 
Eq.~\eqref{FR_0},
$\Omega_0 \to \sqrt{\frac{K}{M}}$ in this case, we get:
\begin{equation}
  \WF_{0} \to \frac \Co{\sqrt[4]{M K}}\Exp{-\int_0^T \frac \varGamma {M} \,\d \tp}.
  \label{sol-lin-osc}
\end{equation}
The last expression coincides with the corresponding result 
\eqref{App1-sol}
for the problem concerning
a free non-stationary oscillation of a mass-spring-damper system
with time-varying parameters (see Appendix~\ref{App-1d}).

In the exceptional cases, where 
\begin{equation}
\Lambda=\const\neq0
\end{equation}
the solution can be found in the form of Eq.~\eqref{sol-Lambda-is-constant}, 
depending on
current values of the system parameters $\bPsi(\pp)$ and the slow time $T$. 
We can underline at least the following such cases:
\begin{enumerate}       
\item Considered in detail in
Appendix~\ref{App-1d} limiting case 
\eqref{case-no-co} of an isolated mass-spring-damper system with
$M=\const$, $\GA=\const$, where $K$ is variable quantity;
\item Limiting case 
\eqref{rho0}
with
$M=\const$, $\GA=\const\neq0$, $\gamma=0$,
where $K$, $k$, $\mathcal T$ are variable quantities, see Eq.~\eqref{WW_final};
\item The case $M=0$, $K=\const$, $\GA=\const\neq0$, 
$\mathcal T=\const$, $\rho=\const$, $\gamma=0$, where $k$ and $v$ are variable
quantities, see Eq.~\eqref{WW_final};
\item The case  $M=0$, $\GA=0$, $\rho=\const$, $\gamma=\const\neq0$, where
$K$,$\mathcal T$, $k$, $v$ are variable quantities, see Eq.~\eqref{WW_final}.
\end{enumerate}

Combining the obtained solution with the complexly conjugated one following to 
Eqs.~\eqref{ansatz-u}--\eqref{ansatz-u-ext}, we get the asymptotic solution
for $\mathcal U$ in the real form:
\begin{equation}
\mathcal U=
2|\Co|\mathcal A(T)\exp\big(-\mathcal
D(T)\big)\cos\left(\int_0^T\Omega_0(\tp)\,\d\tp-\arg \Co\right).
\label{U-expr}
\end{equation}

\begin{remark}
\label{remark-matching}
The unknown complex constant $\Co$ should be found by the matching at
$\xi=0$ and $\tau=0$
the expression 
\eqref{U-expr}
for $\mathcal U$
with results 
\eqref{eq:solution_inhomogeneous_stationary_problem}
obtained by
the method of stationary phase applied to the same system with $\epsilon=0$
and constant parameters, which equal the corresponding initial values. The
procedure is completely analogous to the one presented in Sect.~4.3 of \cite{Gavrilov2022jsv}.
The result is
\begin{equation}
\begin{gathered}
|\Co| = 
\left.
\frac{\left|\F{p}({\O_{0}})\right|}2
\sqrt{\frac{\sqrt{\strut k\TF - k\rho\v^2-\TF\rho\O_{0}^2}}
{\O_{0} M\sqrt{k\TF-k\rho\v^2-\TF\rho\O_{0}^2} +
  \TF\rho\O_{0}}}
\,\,\right|_{T=0}
,
\\
\arg \Co=\arg \F{p}({\O_{0}})-\frac\pi2
,
\end{gathered}
\label{Co-value}
\end{equation}
where symbol $\F{p}(\Omega_0)$ denotes the value of the Fourier transform for the
loading $p(\t)$ calculated at the frequency~$\Omega_0$.
\end{remark}

\selectlanguage{english}

\section{Analysis of previous studies}
\label{sect-previous}
In this section, we discuss in the chronological order the basic results 
of all our previous studies, which
deal with the particular cases of 
Eqs.~\eqref{eq2m}, \eqref{eq1m}
with a single time-varying parameter. 
We show that the earlier analytic results follow from universal formulae
\eqref{final-evolution}, 
\eqref{WW_2} and discuss the method of solving of the first approximation
equation, which was applied in every study. 

For all problems considered in this section we assume the absence of
dissipation, i.e, 
Eq.~\eqref{nondiss-ass} is fulfilled.
All coefficients
for the particular forms of Eq.~\eqref{eq1m} were assumed to be functions of
time $\tau$ only (but not $\xi$), i.e.,
\begin{equation}
\TF'\equiv0, 
\qquad
\rho'\equiv0,
\qquad
k'\equiv0.
\end{equation}
Recall, 
see Remark~\ref{remark-one-parametric},
that in such a case, the right-hand side of the first approximation equation can
always be represented in the form of the exact derivative. Therefore, the
solution is definitely can be obtained in the form of a function depending on
the current values of the system parameters. 
At the same time, while considering different
cases with a single time-varying parameter,
we supposed that it would be rather
impossible to obtain
the result in the form of a function in the case of multiple time-varying parameters, 
since, in the latter case, conditions \eqref{PD2} are not
satisfied automatically any more.

\subsection{A non-uniformly moving inertial load}
\label{sect-PPM}
In \cite{gavrilov2002etm}, the suggested asymptotic approach 
was first time applied to describe a localized
oscillation in a coupled discrete-continuum system
with time-varying properties. Namely, an oscillation of an
inertial load, non-uniformly moving along an infinite string on the Winkler
foundation, was considered. 
To obtain the equations, which coincide with the ones considered in
\cite{gavrilov2002etm} one should formally put in Eqs.~\eqref{eq2m},
\eqref{eq1m}
\begin{equation}
K=0,\qquad \TF=1,\qquad \rho=c^{-2}.
\end{equation}
The unique time-varying parameter was the speed $v$, i.e.,
\begin{equation}
\dot M\equiv 0,
\qquad 
\dot \rho\equiv 0,
\qquad 
\dot k\equiv0.
\end{equation}
In the problem under consideration, we can also put, without loss of
generality, that 
\begin{equation}
k=1,
\qquad 
c=1.
\label{sect-PMM-ass3}
\end{equation}
{This assumption is used in what follows in Sect.~\ref{sect-PPM} for the aim of
simplicity.}

Though the model of a string on the Winkler foundation is widely used in
engineering,
and such a problem has many important applications, before \cite{gavrilov2002etm} 
there was no
approximate solution of a simple structure valid for a non-uniform regime 
of an inertial load motion. 
Indeed, this is an essentially non-stationary problem, which takes into account
both an inertial character of the load \cite{kaplunov1986torsional} and a
non-uniform regime of motion \cite{kaplunov1986vibrations,Gavrilov1999jsv}.
The free oscillation was assumed
in the form
of the single-mode ansatz corresponding to the trapped mode frequency. In this way,
the problem describing the localized oscillation in a distributed
discrete-continuum system with time-varying
parameters was first time reduced to the 
first approximation for the amplitude
in the form of an ODE. 
The procedure of solving of the equation is quite
complicated, since expressions 
\eqref{B} for $B$ and \eqref{S}
for $S$, respectively, were substituted into
the right-hand side of the equation at a very early stage of solving.
Finally, the equation had transformed into the form 
of Eq.~\eqref{gen-K-M-1st-psi-1}
wherein 
\begin{equation}
\Psi_1=v,\qquad
\Psi_2=\Omega_0,\qquad
\Psi_3=v^2,\qquad
\Psi_4=M^2\Omega_0^2+2,\qquad
\Psi_5=1+M^2k(1-v^2)
\end{equation}
and integrated according to Eq.~\eqref{1P-sol}.
Finally, the solution for the amplitude was expressed in the 
form of a product
of eleven integrals.\footnote{The coefficients $\Phi_i$ were sums
of several expressions.}
After simplification, the result was obtained in the form of the
simple algebraic expression:
\begin{equation}
\mathcal W_0=\Co\sqrt{\frac{1-v^2}{\Omega_0(M^2\Omega_0^2+2)}},
\label{W-old}
\end{equation}
see Eq.~(5.15) in \cite{gavrilov2002etm}.
The expression obtained in such a way describes a free localized oscillation 
and involves an unknown constant $C$ {(the initial complex amplitude)}.
The external force $p(\tau)$ was taken in the form of the 
weight of the load, i.e., a suddenly applied constant force. 
Finally, the solution was taken in the
form called the multi-frequency ansatz 
\cite{Gavrilov2022jsv},\footnote{Or, better to say, the multi-modes ansatz.} which takes into account both free and forced
components of the motion. 
To find the
constant $C$, the solution was matched with the results obtained by 
the method of stationary phase applied to the same system with
constant parameters, which are equal to the corresponding initial values.
The constructed analytic solution was verified by numerics based on solving a
Volterra integral equation of the second kind for the internal force $P(\tau)$. 
An excellent
agreement with results of the analytic approach was demonstrated for the case of a
uniformly accelerated motion of the load, 
excepting the case when the trapped mode
frequency approaches the cut-off frequency $\Omega_\ast$.
Nevertheless, later, in \cite{Gavrilov2022jsv}, we showed that formula 
\eqref{W-old} is erroneous, and clearly
identified the error in calculations.
This formula cannot be obtained as a particular case of general
formulae 
\eqref{final-evolution},
\eqref{WW_2}. The results given by the correct formula derived later
in \cite{Gavrilov2022jsv}, see also 
Sect.~\ref{Sect-Gavrilov2022jsv}, are very close 
to ones given by 
\eqref{W-old} and differ considerably only for large enough mass $M$ (see
Sect.~5.2.2 of \cite{Gavrilov2022jsv}).

%

\subsection{An inclusion of a time-varying mass}
In \cite{indeitsev2016evolution}, the particular case 
where 
\begin{equation}
K=0, \qquad \TF=1,\qquad \rho=c^{-2},\qquad  v=0
\label{M(T)-ass1}
\end{equation}
was considered.
The unique time-varying parameter was the mass $M$, i.e.,
\begin{equation}
\dot \rho\equiv 0,
\qquad 
\dot k\equiv0.
\end{equation}
In the problem under consideration we again,
for the aim of simplicity,
can accept assumptions
 \eqref{sect-PMM-ass3}.

Again, 
{expressions} \eqref{B} for $B$ and \eqref{S} for $S$  
were substituted into
the first approximation at a very early stage of solving.
This enough lengthy equation was reduced to one having the
structure of
Eq.~\eqref{gen-K-M-1st-psi-1},
wherein 
\begin{equation}
\Psi_1=m,\qquad
\Psi_2=\Omega_0^2
\end{equation}
and then integrated.\footnote{Note that
\cite{indeitsev2016evolution} is a short communication paper. The details how
we 
reduced the first approximation equation to the form of
Eq.~\eqref{gen-K-M-1st-psi-1}
are not
presented there.}
The obtained expression  
for the amplitude has the form:
\begin{equation}
\mathcal W_0=\Co\frac{\sqrt 2 \sqrt[4]{M}\sqrt[8]{z-2}}{\sqrt{z}\sqrt[8]{z+2}},
\label{W0-DAN}
\end{equation}
see Eq.~(46) in \cite{indeitsev2016evolution}.
Here,
$C$ is an arbitrary constant,
\begin{gather}
z=M^2\Omega^2_0+2,
\label{z}
\\
\Omega_0= \sqrt{\frac{2(\sqrt{1+M^2}-1)}{M^2}}
\label{freq-DAN}
\end{gather}
is the
root 
of frequency equation~\eqref{fr_eq} wherein  
identities 
\eqref{M(T)-ass1} 
and
\eqref{sect-PMM-ass3} 
are taken into account:
\begin{equation}
M\Omega^2=2\sqrt{1-\Omega^2}.
\label{fr_eq-2}
\end{equation}

 Due to Eq.~\eqref{fr_eq-2}, one gets
\begin{equation}
z=M^2\Omega^2_0+\frac{M\Omega^2_0}{\sqrt{1-\Omega^2_0}}
=\frac{M\Omega^2_0(M\sqrt{1-\Omega^2_0}+1)}{\sqrt{1-\Omega^2_0}}.
\label{z-another}
\end{equation}
Taking into account Eqs.~\eqref{z},\eqref{z-another}, one can show that
Eq.~\eqref{W0-DAN} can be transformed as follows:
\begin{equation}
\mathcal W_0
=
\Co\sqrt
2\sqrt{{\frac{\sqrt{1-\Omega^2_0}}{\Omega_0\big(1+M\sqrt{1-\Omega^2_0}\big)}}}
\,
\frac{1}{\sqrt[4]{\Omega_0}\sqrt[8]{M^2\Omega^2_0+4}}
\end{equation}
According to Eq.~\eqref{freq-DAN} we get
\begin{equation}
\frac{1}{\sqrt[4]{\Omega_0}\sqrt[8]{M^2\Omega^2_0+4}}=\frac 1{\sqrt[4]2}.
\end{equation}
Hence, 
\begin{equation}
\mathcal W_0=\Tilde{\Co}\sqrt{{\frac{\sqrt{1-\Omega^2_0}}
{\Omega_0\big(1+M\sqrt{1-\Omega^2_0}\big)}}},
\label{W0-DAN-1}
\end{equation}
where $\Tilde{\Co}=\sqrt[4]2\Co$.
One can see that Eq.~\eqref{W0-DAN-1} coincides with 
Eq.~\eqref{eq:non_stationar_amplitude} 
wherein Eqs.~\eqref{nondiss-ass},
\eqref{sect-PMM-ass3},
\eqref{M(T)-ass1}
are assumed
(with an 
accuracy to an arbitrary constant multiplier).

The external force $p(\tau)$ was taken in the form 
\begin{equation}
p(\tau)=p_\ast\big(H(\tau)-H(\tau-\tau_0)\big),
\label{finite-step}
\end{equation}
where $p_\ast$, $\tau_0>0$ are constants.
Thus, introducing the multi-modes ansatz was not necessary. 
The analytical solution in the case of exponentially vanishing mass $M$ was verified by numerical 
calculations, based on a Volterra integral equation of the second kind for
$P(\tau)$. 
An excellent
agreement with results of the analytic approach was demonstrated for the case
of an exponentially vanishing mass $M$,
excepting the case when the trapped mode
frequency approaches the cut-off frequency $\Omega_\ast$, i.e., $M\simeq0$.

\subsection{A string with time-varying tension}
\label{Sect-Gavrilov2019nody}
In \cite{Gavrilov2019nody}, the particular case where

\begin{equation}
K=-2,\qquad M=0,\qquad \TF=c^2,\qquad \rho=1,\qquad k=1,\qquad  v=0
\label{NoDy-ass}
\end{equation}
was considered.
The unique time-varying parameter was the speed of sound $c$.
The motivations for
such a model 
were related to geophysical applications and a new model of a seismic source
suggested by Prof.~D.A.~Indeitsev in private communications (see conference paper 
\cite{gavrilov2016trapped}). 
This paper is the first of our studies in the series, where the partial
differential equation describing the continuum sub-system cannot be reduced to
an equation with constant coefficients. Accordingly, we had some doubts about
the applicability of a single-mode ansatz to approximate non-stationary
free oscillation in a continuum system with time-varying coefficients. 
Indeed, for a PDE with time-varying coefficients we do
not even have the orthogonality conditions, which allow one to separate
oscillations with different frequencies.

%

In this study, we first time considered the case where the existence of the trapped
mode in zeroth order approximation 
was ensured by a discrete spring of a negative stiffness, but not by the
discrete inertial inclusion. Considering the discrete sub-system in the form of a pure spring significantly
simplified the first approximation for the amplitude, which was obtained in
the very short form:
\begin{equation}
\frac{\d \mathcal W_0}{\mathcal W_0}=
-\frac{\d \Omega_0}{2\Omega_0}-\frac {\d\Omega_0^2}{4(1-\Omega_0^2)}-\frac{\d
c}{2 c},
\label{1app-NoDy}
\end{equation}
see Eqs.~(48), (53) in \cite{Gavrilov2019nody}.
Equation \eqref{1app-NoDy}
has the structure of 
Eq.~\eqref{gen-K-M-1st-psi-1} wherein 
\begin{equation}
\Psi_1=\Omega_0,
\qquad
\Psi_2=\Omega_0^2,
\qquad
\Psi_3=c.
\end{equation}
This results in 
\begin{equation}
\WF_0=\Co\frac{\sqrt[4]{1-\Omega^2_0}}{\sqrt{c\Omega_0}},
\label{W0-T(T)}
\end{equation}
see Eq.~(54) in \cite{Gavrilov2019nody}.
Here $\Omega_0$ can be found by the second formula of
\eqref{FR_0}:
\begin{equation}
\Omega_0^2=1-c^{-2},
\label{freq-T(T)}
\end{equation}
see Eq.~(23) in \cite{Gavrilov2019nody}.
Provided that Eqs.~\eqref{nondiss-ass}, \eqref{NoDy-ass}
are true,
it is easy to see that Eq.~\eqref{W0-T(T)} is the particular case of 
Eq.~\eqref{eq:non_stationar_amplitude}.
	
Our previous numeric
approach based on the Volterra integral equations is not applicable for such a
system. Therefore, to verify the solution, we applied the finite difference
method based on a numerical scheme conserving the energy
\cite{strauss1978numerical,donninger2011numerical}.
Various laws for $c(\epsilon t)$ were considered, in particular a
uniform one, as well as a 
periodic non-monotonous one. The external loading was taken in the form of a finite step function
\eqref{finite-step} or an exponentially vanishing loading.
An excellent agreement with analytic results was shown. It was demonstrated that
Liouville-Green approximation \eqref{L-G} for a single degree of freedom system 
can be clearly distinguished as a wrong solution compared to 
Eq.~\eqref{W0-T(T)} (Fig.~5 in \cite{Gavrilov2019nody}).

We considered such a problem as the first step in the 
investigation of the problem for a Bernoulli-Euler beam with a defect compressed by a
time-varying force, where the trapped modes also can be observed
\cite{gavrilov2016trapped,indeitsev2015localization}.
Though the simplest model problem for
an uncompressed beam with time-varying parameters was solved in
\cite{Shishkina2019jsv}, the problem for a compressed beam remains 
unsolved despite our several attempts. 
For now, we realize that the last problem is much more
difficult, than the reader can think looking through
\cite{Shishkina2019jsv}.

\subsection{A discrete mass-spring sub-system of time-varying stiffness}
In \cite{Gavrilov2019asm}, the particular case where
\begin{equation}
\TF=1,\qquad \rho=1,\qquad k=1,\qquad v=0
\end{equation}
was considered.
The unique time-varying parameter was the spring stiffness $K$, i.e.,
\begin{equation}
\dot M\equiv0.
\end{equation}


The equation for the first approximation was obtained in the simple form
\begin{equation}
\frac{\d \mathcal W_0}{\mathcal W_0}=
-
\frac{{\d\Omega_0}}{2\Omega_0}
-
\frac{\Omega_0\,\d\Omega_0}{2(1-{\Omega}^2_0)\Big(1+M\SQRTO\Big)}
,
\end{equation}
where $\Omega_0$ is the trapped mode frequency, see Eq.~(62)
in \cite{Gavrilov2019asm}.
The final result coincides
with
Eq.~\eqref{W0-DAN-1}.
To verify the solution numerically, 
we again applied the finite difference method. We have shown that the analytic
and numerical results are in excellent agreement. It was demonstrated that the
results are clearly visually distinguishable with the solution 
\eqref{L-G}, which describes the single degree of freedom system with
time-varying spring. The latter problem is a limiting case of the problem under
consideration as $M\to\infty,\ K\to\infty,\ K/M=\Omega_0^2$.


\subsection{A non-uniformly moving  discrete mass-spring sub-system}
\label{Sect-Gavrilov2022jsv}
In \cite{Gavrilov2022jsv}, the particular case where
\begin{equation}
\TF=1,\qquad\rho=1,\qquad k=1
\label{JSV-as1}
\end{equation}
was considered.
The unique time-varying parameter was the speed $v$, i.e.,
\begin{equation}
\dot K\equiv0,
\qquad 
\dot M\equiv 0.
\label{JSV-as2}
\end{equation}
This problem, being an extension of the problem considered in the first paper
\cite{gavrilov2002etm}, possesses more rich dynamics than the one considered
before.  Indeed, in the latter case the critical speed $v=1$ generally does not coincide
with
the speed $v$ satisfying condition~\eqref{loc-cond}, at which
the trapped mode disappears. The case $K<0$ was also
included into consideration. Besides a free localized oscillation, the
forced oscillation caused by a class of loadings, which 
can be represented as a superposition of a pulse loading, which
acts during some time, and a number of harmonics with slowly time-varying
frequencies and  amplitudes is considered, see Eqs.~\eqref{p-def}--\eqref{non-resonant}.
The solution was looked for in the form of
a superposition of a free oscillation with frequency $\Omega_0$ and  forced
oscillations with frequencies $\Omega_i,\ i=\overline{1,N}$,  i.e., in the form
of the multi-modes ansatz. The forced oscillations can be found by zero
order approximation in the framework of the method of multiple scales.
Considering the equation for the first approximation in the form 
of Eq.~\eqref{before-solving} where 
Eqs.~\eqref{nondiss-ass},
\eqref{JSV-as1}, \eqref{JSV-as2} are taken into account, 
and {expressions} \eqref{B} for $B$ and \eqref{S} for $S$ are not substituted,
we
did not use the approach discussed in 
Remark~\ref{remark-one-parametric}, which was applied in all previous papers.
Instead, 
new variables 
$\bPsi$ \eqref{Psi-hack} were introduced. Utilizing of variables \eqref{Psi-hack}
allowed us to  
reduce 
the equation for the first approximation
to the form of Eq.~\eqref{exact-d} due to unobvious relationship
\eqref{relationship} between the system parameters.
Thus, the expression for the amplitude was obtained in
the form of Eqs.~\eqref{final-evolution},
\eqref{WW_2}
wherein Eqs.~\eqref{nondiss-ass}, 
\eqref{JSV-as1} are taken into account.
The obtained result corrects the error
introduced in \cite{gavrilov2002etm}, see Sect.~\ref{sect-PPM}.

The constructed analytic solution was verified by numerics based on solving a
Volterra integral equation of the second kind for the internal force $P(\tau)$. 
An excellent
agreement with results of analytic approach was demonstrated for the cases of
a pure free and forced oscillation.
%
%
%

\subsection{Discussion}
\label{sect-disc}

After the publication of paper \cite{Gavrilov2022jsv}, we have realized that the
procedure suggested there allows one to obtain the solution for the amplitude
in the form of a function for an extended non-dissipative problem,
where all the parameters
are assumed to be time-varying quantities, which is considered in this paper.
The results of this paper is an
essential 
generalization of \cite{Gavrilov2022jsv}, which becomes possible due to
a new approach based on discovered in 
\cite{Gavrilov2022jsv} 
lucky choice \eqref{Psi-hack}
of variables $\bPsi$.
{The fact that 
for the multi-parametric case of 
Eq.~\eqref{WW_1}
conditions 
\eqref{PD2} are fulfilled, and the logarithmic structure of the right-hand
side is still observed (see Remark~\ref{remark-log}), is surprising
for us. Apparently, from the mathematical point of view, it follows from some
unobvious properties of governing equations \eqref{eq2m}, \eqref{eq1m}.}

The previous studies
\cite{Gavrilov2022jsv,Gavrilov2019nody,Gavrilov2019asm,indeitsev2016evolution,gavrilov2002etm}
still keep the importance after this paper, in particular,
due to numerical work done there.
Indeed, in previous studies the practical applicability of the formal
asymptotics
to a wide class of systems with various single time-varying parameters,
various regimes of time-varying, and various loadings was demonstrated in
the case 
when the hodograph of the tuple $\pp(0,\tau)$ introduced by
Eqs.~\eqref{pP}
\eqref{s-parameters} does not approach the boundaries of the localization
domain $\LL$. The suggested approach allows us to investigate
analytically the practically important problems,
where non-uniformly moving mass-spring system is
under consideration \cite{Gavrilov2022jsv,gavrilov2002etm}. The same approach
was also applied \cite{Gavrilov-2006-trans} to find the law of motion for a discrete moving load 
subjected to a drug force and interacting with the continuum sub-system by so-called
``wave pressure force'', see 
\cite{rayleigh1902xxxiv,Nicolai1912-eng,nicolai1925xix,Gavrilov2016nody,havelock1924lxx,Gavrilov(ActaMech),Slepyan2017,Slepyan2017a,Denisov2012,Brillouin1925,Vesnitski1983,Andrianov1993},
or ``external configurational force''
\cite{Cherepanov1985,gurtin2000cfb}.

The formal asymptotics
generally breaks for a system with parameters such that the trapped mode frequency
becomes close to the cut-off frequency, i.e., when $\pp(0,\tau)$ 
approaches for certain values of $\tau$ the boundary of
$\LL$ defined by Eq.~\eqref{loc-cond}.
In another limiting case, where the trapped
mode frequency becomes zero, i.e., when
$\pp(0,\tau)$ approaches the boundary of $\LL$ defined by Eq.~\eqref{stab-cond},
the formal asymptotics usually describes the solution quite well until the loss of stability, which
leads to a localized buckling. The cases when 
$\pp(0,\tau)$ approaches the boundaries \eqref{loc-cond} or \eqref{stab-cond}
are discussed in more detail in what follows in the present paper, see 
Sect.~\ref{As-breaks}.

\section{Numerics}
As well, as we have already discussed in the Introduction, the obtained asymptotics 
is so-called formal asymptotics, i.e.
it satisfies the corresponding equations and initial conditions up to the
certain asymptotic accuracy. 
Moreover,
when constructing the asymptotics, a number of not well grounded 
assumptions were used. The most voluntaristic of them is the possibility to
represent a non-stationary oscillation of a system 
with time-varying parameters in the form of a single-mode ansatz, which
corresponds to ``a~pure localized oscillation'' at a given frequency. 
Dealing with PDE with time-varying coefficients, we do
not even have the orthogonality conditions, which allow one to separate
oscillations with different frequencies. Another question is the validity
of the matching procedure for two asymptotics, which were obtained by
entirely different approaches. 
Thus, in our opinion, the analytic results should be at least justified by
independent numerical calculations.

As it is mentioned in Sect.~\ref{sect-previous}, there are two basic numeric
approaches, which can be applied.
The first approach is related with a
Volterra integral equation of the second kind, 
to which the problem under consideration can be
reduced in some particular cases 
\cite{Gavrilov2022jsv,indeitsev2016evolution,gavrilov2002etm}.
A more general alternative method is
based on
the finite difference schemes used in
\cite{Gavrilov2019nody,Gavrilov2019asm}. The various particular cases, where the
only one parameter varies in time, were considered in the previous studies.
To verify the constructed single-mode asymptotics, in this paper, we consider only two particular cases. The first
one is the case of a system with two simultaneously time-varying
parameters (Sect.~\ref{Sect-two-pa}).
The second one is related with the dissipative case
(Sect.~\ref{Sect-num-diss}). Both cases require 
the second approach. 

\begin{remark}	
\label{remark-sym}
We restrict
ourselves with the symmetric with respect to $\xi$ problems, since only for such
problems we have the debugged code for now. 
The non-symmetric
cases, e.g., the cases where $v\neq0$ or some parameters of the continuum
sub-system are spatially varying functions, require considerable
additional programming work, which is beyond the scope of this paper.
\end{remark}

Additionally, in Sect.~\ref{As-breaks} we consider several examples of
systems, where the hodograph of the tuple $\pp(0,\tau)$ introduced by 
Eq.~\eqref{s-parameters} approaches the boundary of the localization domain
$\LL$. In particular, in such a way, we show that the good practical
applicability of 
the constructed
single-mode asymptotics is related with the phenomenon of the anti-localization
of non-stationary linear waves \cite{Gavrilov2024,Shishkina2023cmat,Shishkina2023jsv,Gavrilov2023DD}.

The external force $p(\tau)$ for numerics is taken in the form of Eq.~\eqref{finite-step} 
wherein $p_\ast=1/\tau_0$. Thus, $p(\tau)\to\delta(\tau)$ in the weak
sense. The corresponding asymptotics is taken for the case
$p(\tau)=\delta(\tau)$.
In all following examples, the value of the small parameter is 
\begin{equation}
\epsilon=0.01,
\end{equation}
and 
\begin{equation}
\tau_0=0.01.
\end{equation}
\subsection{Two time-varying parameters in a non-dissipative system} 
\label{Sect-two-pa}
Consider the particular case where equations 
\begin{equation}
M=1,\qquad \rho=1,\qquad k=1,\qquad v=0
\label{2p-as}
\end{equation}
and Eq.~\eqref{nondiss-ass} are fulfilled. The time-varying parameters are $\mathcal T$
and $K$:
\begin{gather}
    \mathcal T(T) = 1.0-0.9\sin(0.4 T),
\label{all-time1}
\\
    K(T)=-0.6-0.8\sin(1.5T).
\label{all-time2}
\end{gather}%
The plots of $\mathcal T$, $K$ and the corresponding
value of the trapped mode frequency $\Omega_0$ are presented 
in Fig.~\ref{all-time.pdf}.
In Fig.~\ref{u-time.pdf} we compare the corresponding asymptotic and numerical
solutions.
One can see that the solutions
are in an excellent agreement. 
\begin{figure}[p]  
\centering\includegraphics[width=0.55\textwidth]{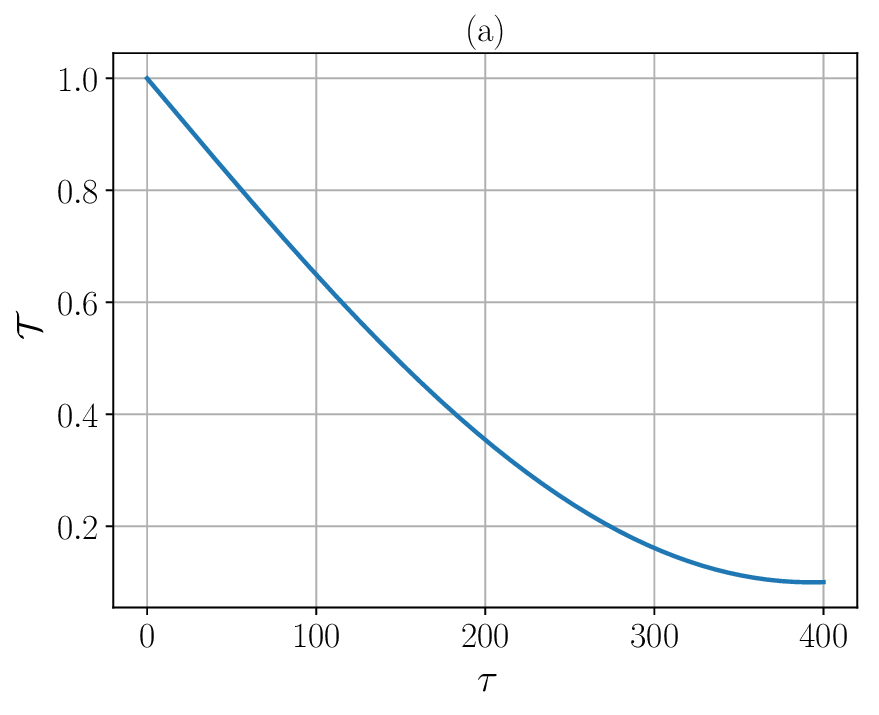}
\centering\includegraphics[width=0.55\textwidth]{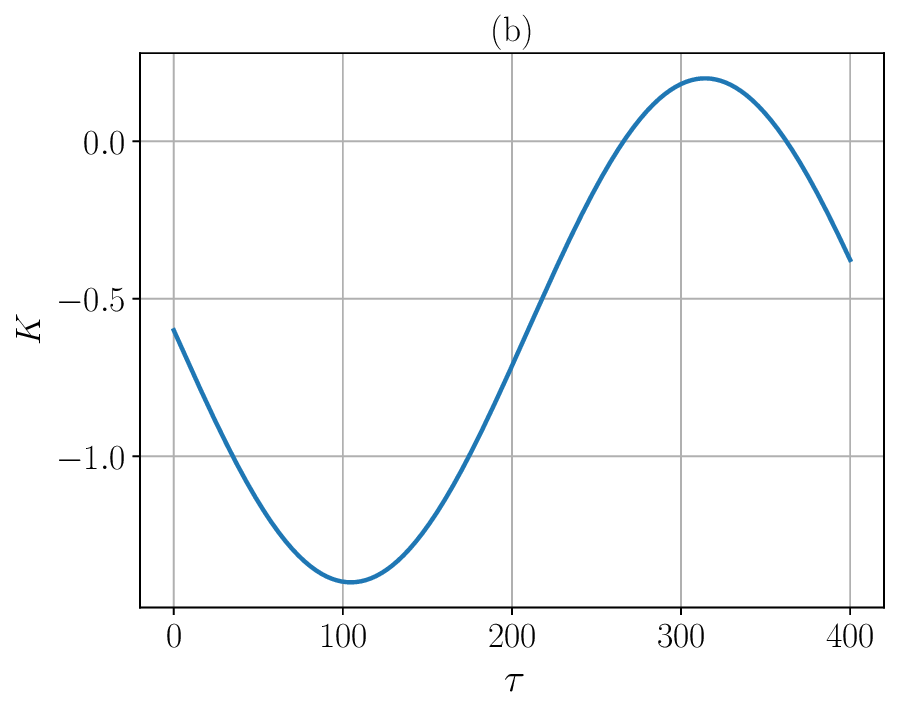}
\centering\includegraphics[width=0.55\textwidth]{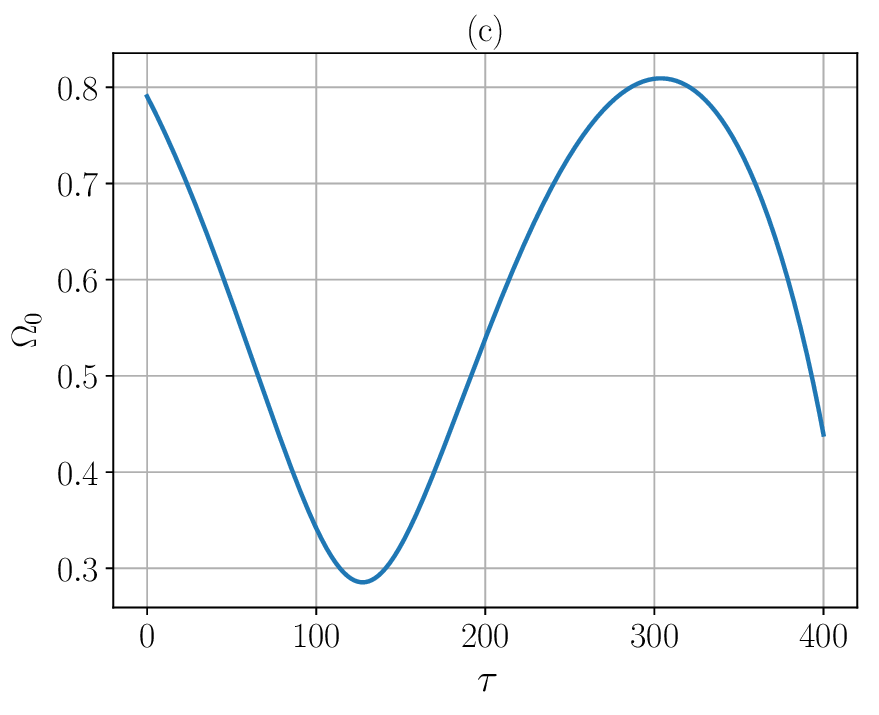}
\caption{Given time-varying parameters (a) $\mathcal T$, (b) $K$, and (c) the trapped
mode frequency $\Omega_0$  for the case when parameters
are taken according Eqs.~\eqref{2p-as},
\eqref{all-time1}, \eqref{all-time2}}
\label{all-time.pdf}
\end{figure}

\begin{figure}[htb]  
\centering\includegraphics[width=0.65\textwidth]{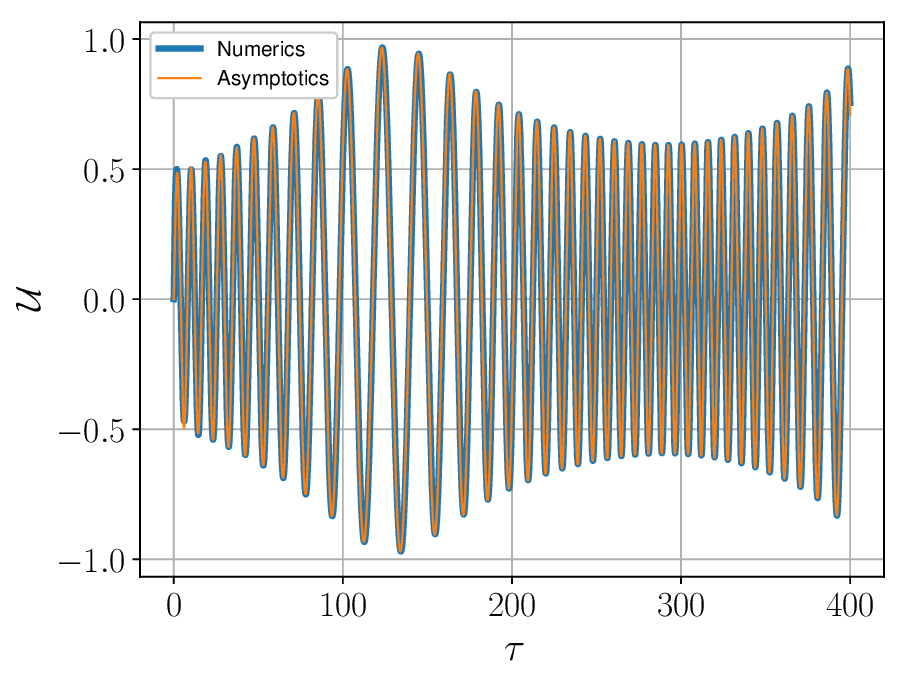}
\caption{Oscillation of the discrete sub-system for the case when parameters
are taken according Eqs.~\eqref{nondiss-ass}, 
\eqref{2p-as}, \eqref{all-time1}, \eqref{all-time2}}
\label{u-time.pdf}
\end{figure}


\subsection{A dissipative system}
\label{Sect-num-diss}
Consider the particular case where equations 
\begin{equation}
M=1,\qquad\mathcal T=1,\qquad\rho=1,\qquad k=1,\qquad  v=0,
\qquad \varGamma=1,\qquad \gamma=0
\label{Gamma-as}
\end{equation}
are satisfied.
The only time-varying parameter is $K$:
\begin{equation}
K(T)=K_0+K_1 T.
\label{K-linear}
\end{equation}
Note that $K$ should satisfy restriction
\begin{equation}
-2<K<1,
\label{K-cond-1}
\end{equation}
which follows from inequalities~\eqref{stab-cond}, \eqref{loc-cond}, 
wherein 
Eq.~\eqref{Gamma-as} is used.
For squared frequency $\Omega_0$ according to Eq.~\eqref{FR_0} we have
\begin{equation}
\Omega^2_0=K-2+2\sqrt{2-K}.
\end{equation}
Taking into account the above expression, from Eqs.~\eqref{final-evolution}, 
\eqref{WW_final} one
gets:
\begin{equation}
\label{WW_final-part}
\WF_{0} = \Co
\sqrt{
\frac{\sqrt{2-K}-1}
{
\sqrt{2-K}\sqrt{2\sqrt{2-K}-(2-K)}
}
}
\, \Exp{-\int_0^T \left(1-\frac{1}{\sqrt{2-K(\tp)}}\right)\, \d \tp}.
\end{equation}
Substituting Eq.~\eqref{K-linear} into the expression for $\WF_{0}$, we
derive:
\begin{equation}
\WF_{0} = \Co
\sqrt{
\frac{\sqrt{2-K}-1}
{
\sqrt{2-K}\sqrt{2\sqrt{2-K}-(2-K)}
}
}
\,
\Exp{-T-\frac{2}{K_1}\left(\sqrt{2-K_0-K_1T}-\sqrt{2-K_0}\right)}.
\label{WW_final-part-1}
\end{equation}

\begin{remark}  
Note that for $K_1 \to 0$ one has:
\begin{equation}
\lim_{K_1\to0}\left(-T-\frac{2}{K_1}\left(\sqrt{2-K_0-K_1T}-\sqrt{2-K_0}\right)\right)
=
-T\left(1-\frac {1}{\sqrt{2-K_0}}\right).
\end{equation}
Thus, the right-hand side of 
\eqref{WW_final-part-1} is properly defined for $K_1=0$.
\end{remark}

In Figs.~\ref{u-time-Gamma.pdf}, \ref{u-time-Gamma-inc.pdf} we show 
the plots of the stiffness $K$, the trapped mode frequency $\Omega_0$
and compare asymptotic and numerical solutions for
the case when parameters of the system 
are taken according to Eqs.~\eqref{Gamma-as},
\eqref{K-linear} wherein $K_0=0.5,\ K_1=-1.0$ and $K_0=-1.5,\ K_1=1.0$,
respectively.
One can see that the asymptotic and numerical solutions
are in an excellent agreement, everywhere excepting the case when
{$\Omega_0\simeq \Omega_\ast=1$.}
The analytic solutions, which correspond to the
same system with $\varGamma=0$, are also shown.

\begin{figure}[p]  
\centering\includegraphics[width=0.55\textwidth]{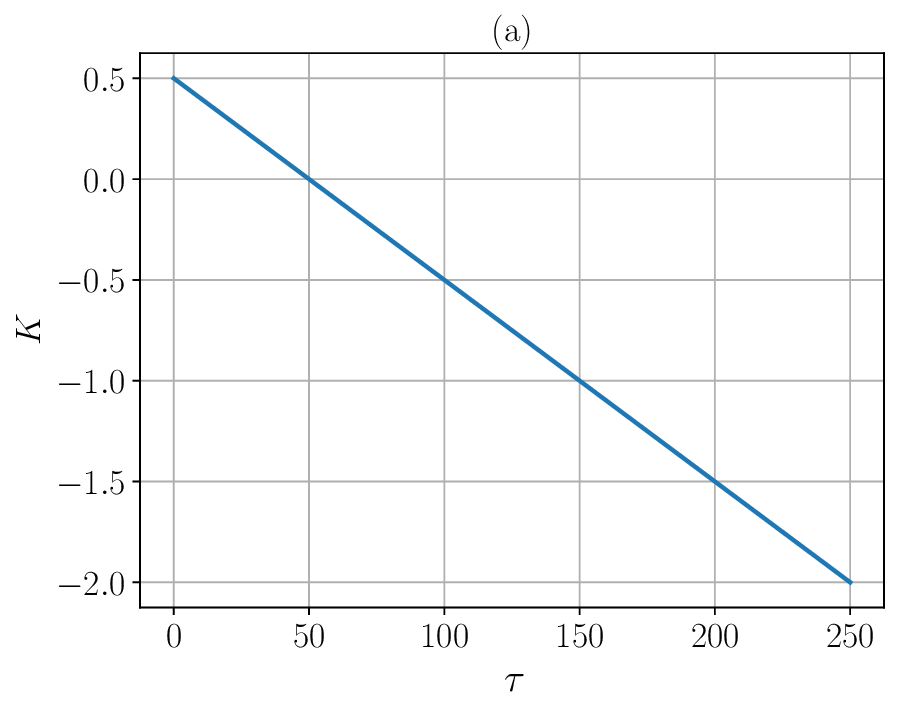}
\centering\includegraphics[width=0.55\textwidth]{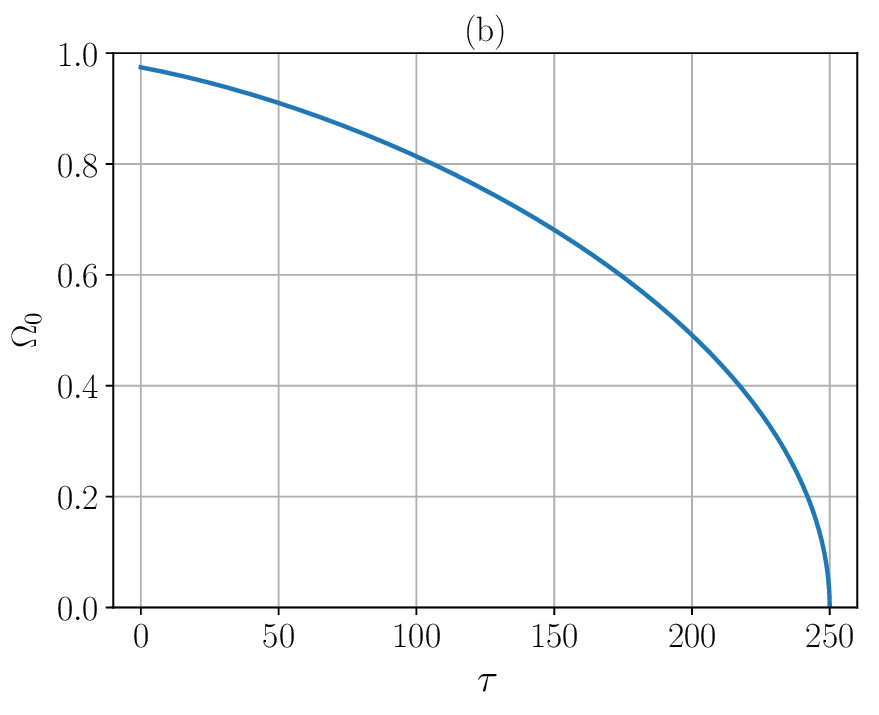}
%
\centering\includegraphics[width=0.55\textwidth]{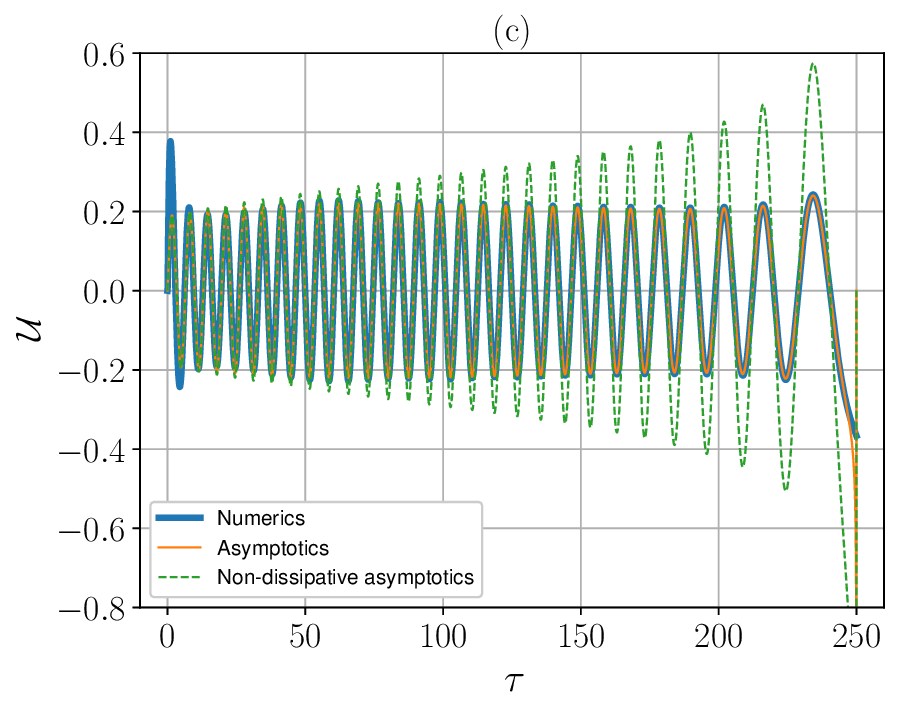}
\caption{
(a) Given time-varying parameter $K$,
(b) the trapped
mode frequency $\Omega_0$, and (c) the displacement $\mathcal U$ of the discrete sub-system for the case when parameters
are taken according Eq.~\eqref{Gamma-as}, \eqref{K-linear} wherein $K_0=0.5,\
K_1=-1.0$}
\label{u-time-Gamma.pdf}
\end{figure}

\begin{figure}[p]  
\centering\includegraphics[width=0.55\textwidth]{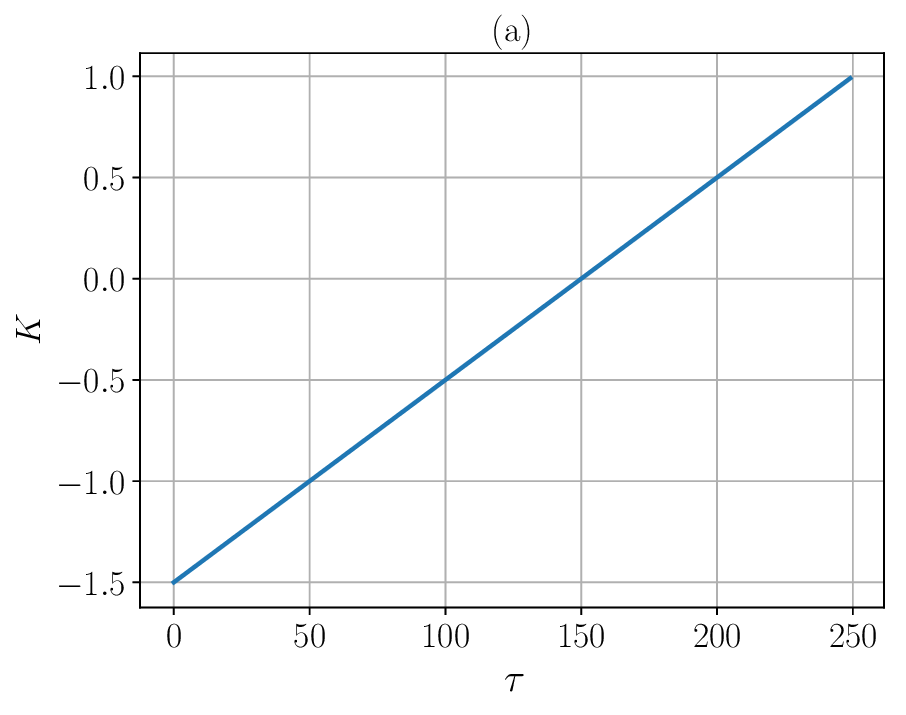}
\centering\includegraphics[width=0.55\textwidth]{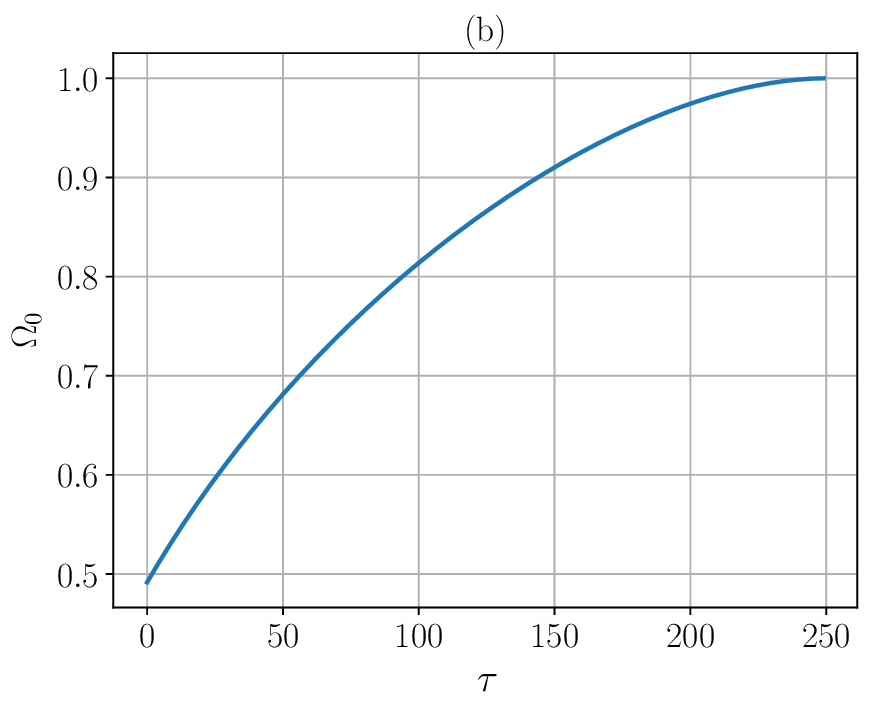}
%
\centering\includegraphics[width=0.55\textwidth]{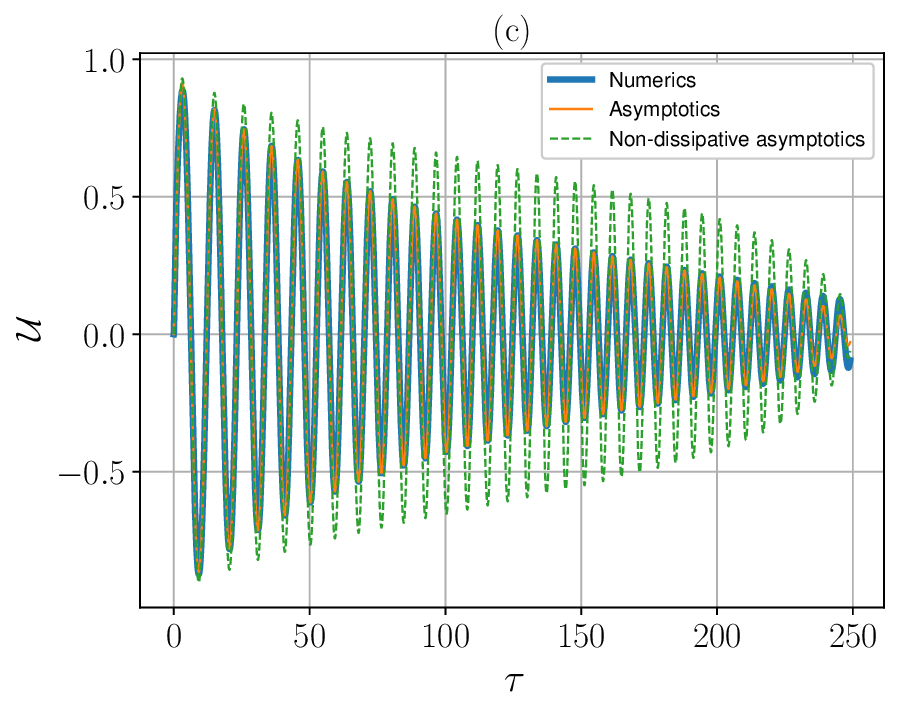}
\caption{(a) Given time-varying parameter $K$,
(b) the trapped
mode frequency $\Omega_0$, and (c) 
the displacement $\mathcal U$ of the discrete sub-system for the case when parameters
are taken according Eq.~\eqref{Gamma-as}, \eqref{K-linear} wherein $K_0=-1.5,\
K_1=1.0$}
\label{u-time-Gamma-inc.pdf}
\end{figure}

\subsection{Systems with parameters approaching and crossing the boundaries of the
localization domain}
\label{As-breaks}

Here we consider two examples of the systems where the constructed single-mode
asymptotics is not practically applicable.
These examples are related to
situations, when the hodograph of the tuple $\pp(0,\tau)$
introduced by 
Eq.~\eqref{s-parameters} approaches the boundaries of the localization domain
$\LL$
defined by inequalities 
\eqref{stab-cond}, \eqref{loc-cond}.
Finally, we plan to discuss why the single-mode
approximation is so good for systems with instantaneous values of
parameters lying always far from the
boundaries of the localization domain $\LL$.

The first example corresponds to the case when the trapped mode frequency
reaches an almost zero positive value, 
and then again moves away from zero. This
corresponds to the case when the hodograph of the tuple $\pp(0,\tau)$ 
approaches a neighbourhood of the boundary of the localization domain 
$\LL$ defined by inequality \eqref{stab-cond} and then returns inside $\LL$
without any intersection.
Consider a particular case of such a system, where Eqs.~\eqref{2p-as}
\eqref{all-time1} are fulfilled 
and the relationship
\begin{gather}
    K(T)=-0.6-0.95\sin(1.5T)
\label{all-time2-mod}
\end{gather}
is used instead of 
Eq.~\eqref{all-time2}.
The plots of $K$ and the corresponding
value of the trapped mode frequency $\Omega_0$ are presented 
in Fig.~\ref{all1-time.pdf}(a--b) (the plot of $\mathcal T$ is shown in 
Fig.~\ref{all-time.pdf}(a)). One can see that at $\tau\approx125$ the trapped
mode frequency approaches zero (the minimal value is approximately $0.03$) and
again increases.
In Fig.~\ref{all1-time.pdf}(c) we compare the corresponding asymptotic and
numerical solutions. Looking at the plot of the numerical solution, 
one can observe essential increasing of the oscillation amplitude after
$\tau\approx125$, which is not described by the asymptotic solution. Note that
repeating the numerical calculations with 
decreased both space and time step sizes by factor of 2 yields the same
results, thus, the observed effect, apparently, is
not related to a numerical instability. For larger values of time
($\tau\gtrsim125$), when the parameters are again far enough  from the boundary of
the localization domain, the asymptotic solution again correctly predicts the phase
value and the profile of
the amplitude evolution, but the value of amplitude remains wrong.

\begin{remark}	
It is well known that in the limiting case of the mass-spring system 
\eqref{case-no-co} the special considerations are necessary to obtain the
correct asymptotics in the neighbourhood of points, where $\Omega^2=0$, i.e.,
the turning points, see, e.g., \cite{Jeffreys1925,McHugh1971}.
We do not consider in this paper such subtle effects, and we are not sure that
it could be meaningful in the case of the composite system under consideration.
\end{remark}

\begin{figure}[p]  
\centering\includegraphics[width=0.55\textwidth]{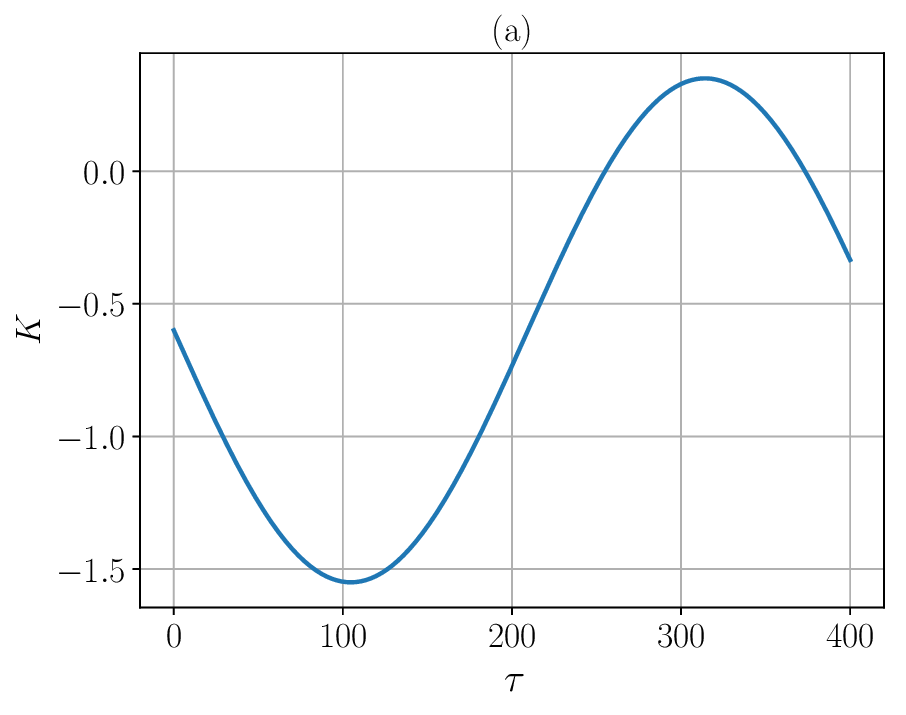}
\centering\includegraphics[width=0.55\textwidth]{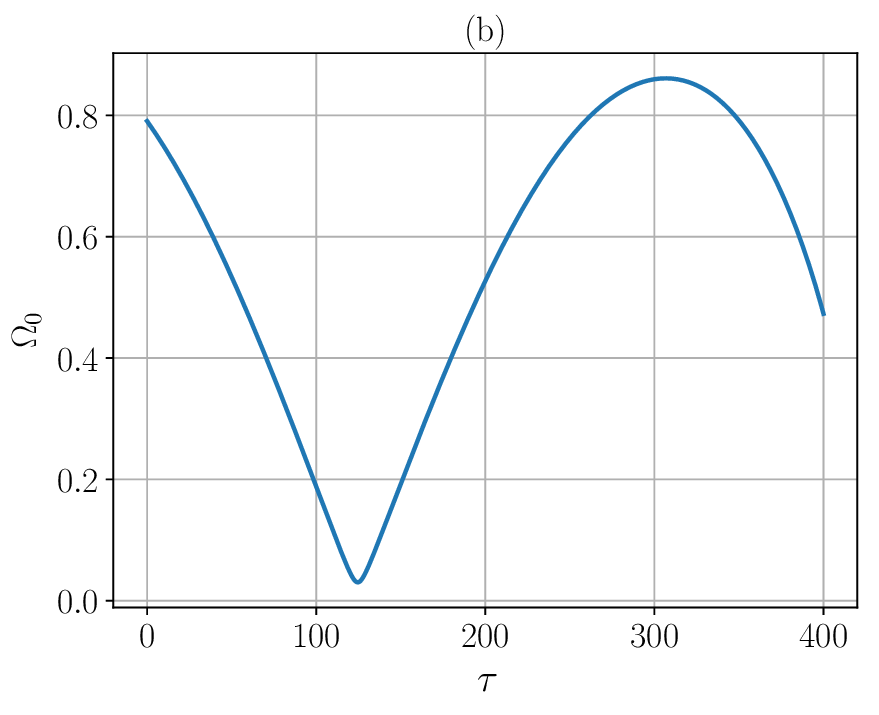}
\centering\includegraphics[width=0.55\textwidth]{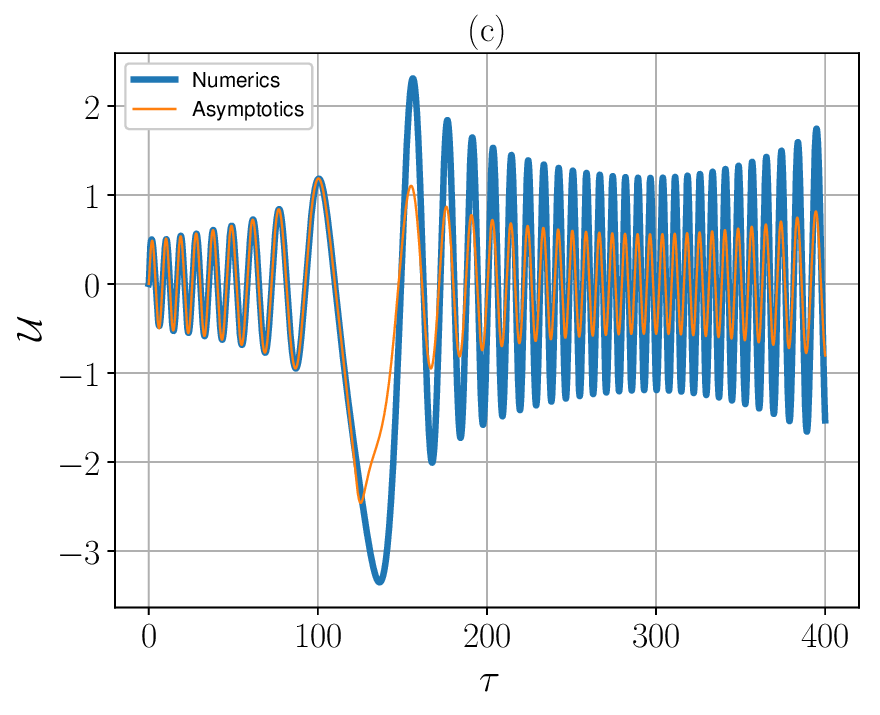}
\caption{(a) Given time-varying parameter $K$, (b) the trapped
mode frequency $\Omega_0$, (c) the displacement $\mathcal U$ of the discrete sub-system 
for the case when parameters
are taken according Eqs.~\eqref{2p-as},
\eqref{all-time1},
\eqref{all-time2-mod}}
\label{all1-time.pdf}
\end{figure}

The second example is more interesting. As we already noted
in Sect.~\ref{sect-disc}, in previous studies 
\cite{Gavrilov2022jsv,Gavrilov2019nody,Gavrilov2019asm,indeitsev2016evolution,gavrilov2002etm}
it was observed that 
the formal single-mode asymptotics generally breaks for a system with parameters such
that the trapped mode frequency becomes close to the cut-off frequency. Let us
consider an example of such a system in more details. Consider the system with
a single time-varying parameter, where 
Eq.~\eqref{nondiss-ass} and equations
\begin{gather}
M=1,\qquad \mathcal T=1,\qquad \rho=1,\qquad k=1,\qquad  v=0,
\label{2p-asT}
\\
K=1-\exp(-3T)
\label{K-exp}
\end{gather}
are fulfilled.
The plots of $K$ and the corresponding
value of the trapped mode frequency $\Omega_0$ are presented 
in Fig.~\ref{all2-time.pdf}(a--b). One can see that for large values of time,
the trapped mode frequency approaches the cut-off frequency $\Omega_\ast=1$,
see 
Eqs.~\eqref{cutoff}, \eqref{FR_0}. Comparing in Fig.~\ref{all2-time.pdf}(c) the corresponding asymptotic and
numerical solutions, one can observe two stages. At the first stage
{($\tau\lesssim 50$)}, asymptotics describes the numerical solution quite well.
At the next stage, the amplitude of the constructed single mode asymptotics,
clearly, approaches zero
{faster} than the corresponding numerical solution.
Namely, according to Eqs.~\eqref{final-evolution},
\eqref{eq:non_stationar_amplitude},
\eqref{U-expr},
\eqref{Co-value}
one has
\begin{equation}
\mathcal W=C_1\Exp{-\frac32\epsilon\t}+o\left(\Exp{-\frac32\epsilon\t}\right),
\quad\t\to\infty;\qquad C_1\approx 0.4.
\end{equation}
At the same time, one can qualitatively estimate the asymptotic decay 
rate for the amplitude of the numerical solution at the second stage as 
\begin{equation}
\mathcal W=\frac{C_2}{\sqrt \t}+o\left(\frac1{\sqrt \t}\right),\quad
\t\to\infty;\qquad C_2=\text{const}\neq0. 
\end{equation}
(see
Fig.~\ref{all2-time.pdf}(c)). It is natural to assume that at the second stage
the system under consideration can be treated as a system with almost constant
parameters lying at the boundary of the localization domain 
where
$\Omega_0\simeq\Omega_\ast$. Actually, in such a system 
with constant parameters, the order of decay is $\t^{-1/2}$ \cite{Shishkina2023jsv}. Thus,
apparently, the approximate description for $\mathcal U$ can be obtained as a matched
asymptotics, which stages correspond to the single-mode approximation for
the system with time-varying parameters and asymptotics for the same system
with constant parameters, respectively. 

\begin{figure}[p]  
\centering\includegraphics[width=0.55\textwidth]{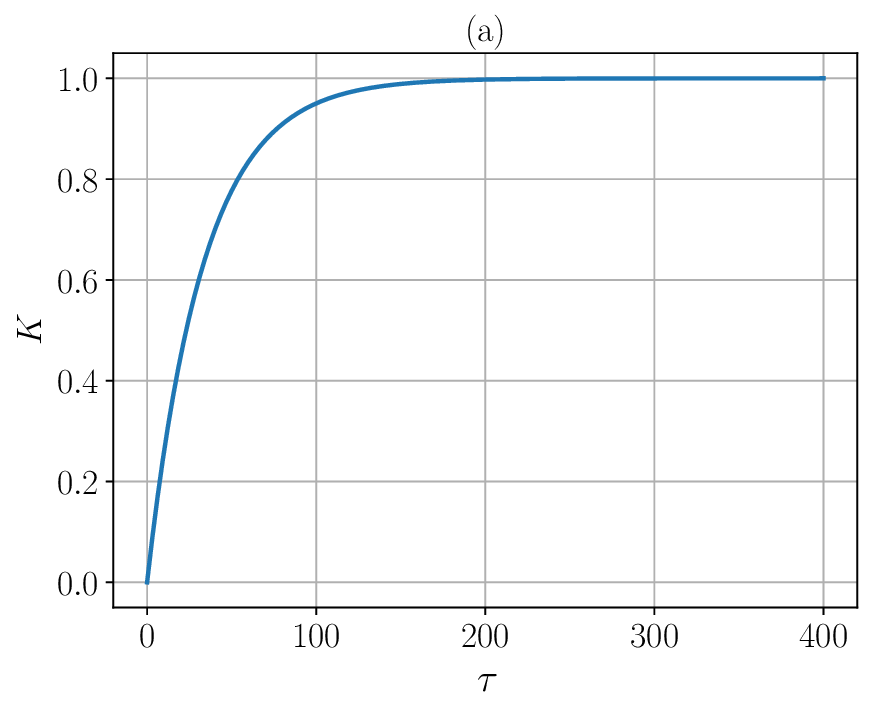}
\centering\includegraphics[width=0.55\textwidth]{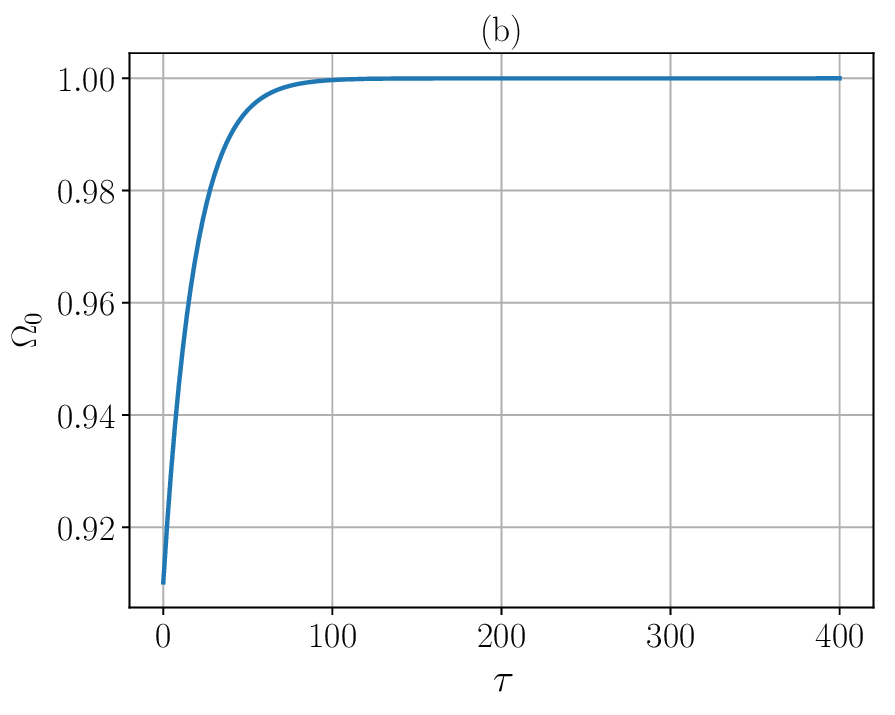}
\centering\includegraphics[width=0.55\textwidth]{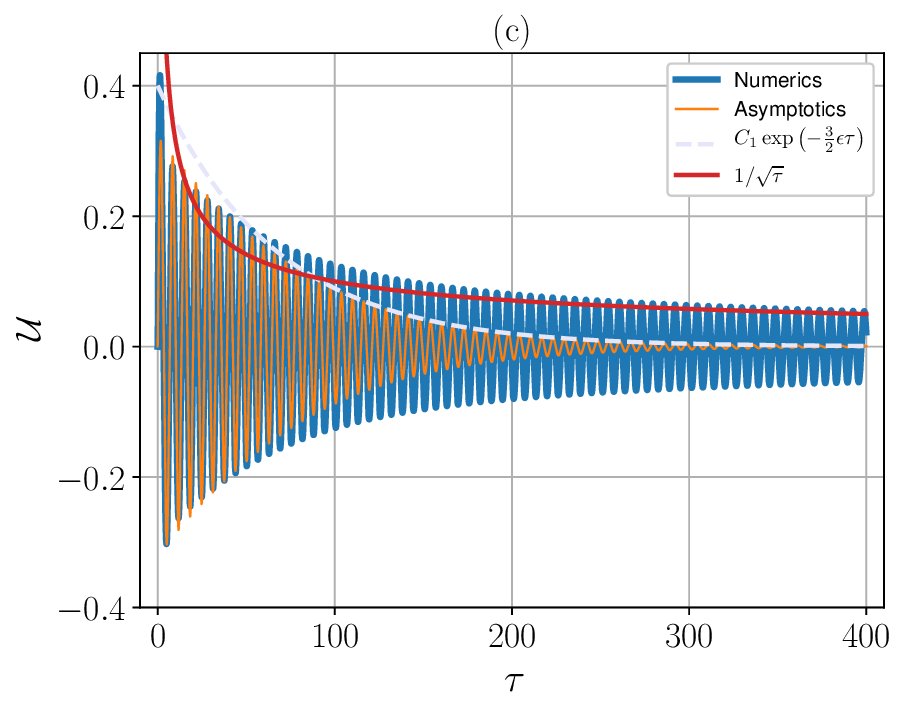}
\caption{(a) Given time-varying parameter $K$, (b) the trapped
mode frequency $\Omega_0$, (c) the displacement $\mathcal U$ of the discrete sub-system 
for the case when parameters
are taken according Eqs.~\eqref{2p-asT},
\eqref{K-exp}}
\label{all2-time.pdf}
\end{figure}
\afterpage{\clearpage}

One can see that in Fig.~\ref{all2-time.pdf}(c) oscillation with
the asymptotic rate of decay $\t^{-1/2}$ is clearly recognizable. So the question
arises why such an oscillation is not recognizable in 
Figs.~\ref{u-time.pdf}, 
\ref{u-time-Gamma.pdf}(b),
\ref{u-time-Gamma-inc.pdf}(b)
and in all corresponding plots obtained in previous papers 
\cite{Gavrilov2022jsv,Gavrilov2019nody,Gavrilov2019asm,indeitsev2016evolution,gavrilov2002etm}.
Why is the single-mode approximation so good inside the localization domain?
Indeed, we expect that in an infinite spatially uniform one-dimensional 
continuum system without
trapped modes, the perturbations decay with the rate $\t^{-1/2}$
\cite{Whitham1999,Slepyan1972}.\footnote{The rate of decay $\t^{-1/2}$ corresponds to the
contribution from a stationary point in the representation of the time-domain
fundamental solution in the form of the inverse Fourier transform.}
Of course, this is true for a spatially uniform pure continuum system. Nevertheless, for continuum
systems with inclusions, the corresponding decaying as  $\t^{-1/2}$ oscillation is also
observed everywhere between the leading wave-fronts excepting a neighbourhood
of a discrete inclusion.
However, the mode decaying
with the rate $\t^{-1/2}$ always has zero amplitude near the inclusion, excepting the 
case, when the system parameters correspond to the boundary of the
localization domain where $\Omega_0\to\Omega_\ast$. The particular case of
such a system is a uniform pure continuum system.
We have called this wave phenomenon the
anti-localization of non-stationary linear waves 
\cite{Gavrilov2024,Gavrilov2023DD,Shishkina2023cmat,Shishkina2023jsv}.
Thus, just the existence of the anti-localization makes the single-mode
asymptotics to be so good approximation for a system with parameters inside
the localization domain far from its boundaries.

On the other hand, for a system {where the hodograph of
$\pp(0,T)$
transversely crosses the boundary of $\LL$ where}
$\Omega_0\to\Omega_\ast$,
we expect fast decaying of the
oscillation amplitude $\mathcal W$. Indeed, due to the anti-localization, the
rate of decay near
the discrete sub-system for the corresponding system with constant parameters 
lying outside $\LL$ is 
$\t^{-3/2}$
\cite{Shishkina2023jsv}.
In Fig.~\ref{all3-time.pdf} we illustrate this fact for the system 
where 
Eqs.~\eqref{nondiss-ass}, \eqref{2p-asT} and 
\begin{equation}
K=0.5+0.5T
\end{equation}
are fulfilled.

\begin{remark}  
For the particular case of the discrete sub-system in the form of pure
inertial inclusion, the result, which shows that  
the decay rate for the vanishing component is
$\t^{-3/2}$ was obtained by Kaplunov in \cite{kaplunov1986torsional}, see also earlier studies
\cite{hemmer1959dynamic,Kashiwamura1962,Mueller1962,Mueller2012,Rubin1963}
concerning infinite discrete systems.  
In paper \cite{Shishkina2023jsv}, we first time have demonstrated that the emergence and the
intensity of the anti-localization are related not with its
non-uniformity itself, but with the position of $\pp$ far enough 
from the boundary of the localization domain where
$\Omega_0\to\Omega_\ast$. 
\end{remark}

\begin{figure}[p]  
\centering\includegraphics[width=0.55\textwidth]{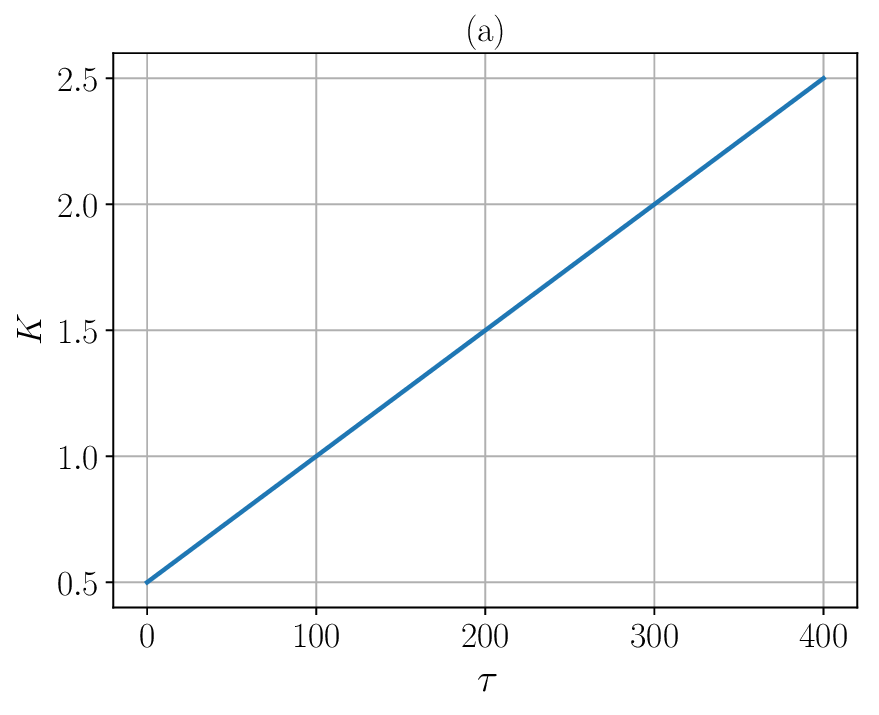}
\centering\includegraphics[width=0.55\textwidth]{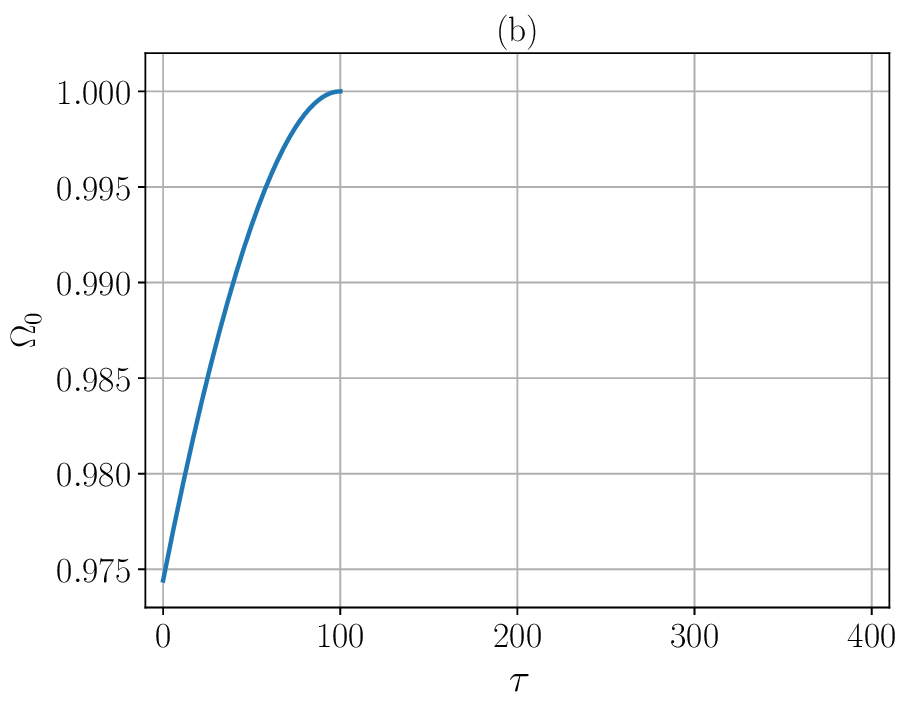}
\centering\includegraphics[width=0.55\textwidth]{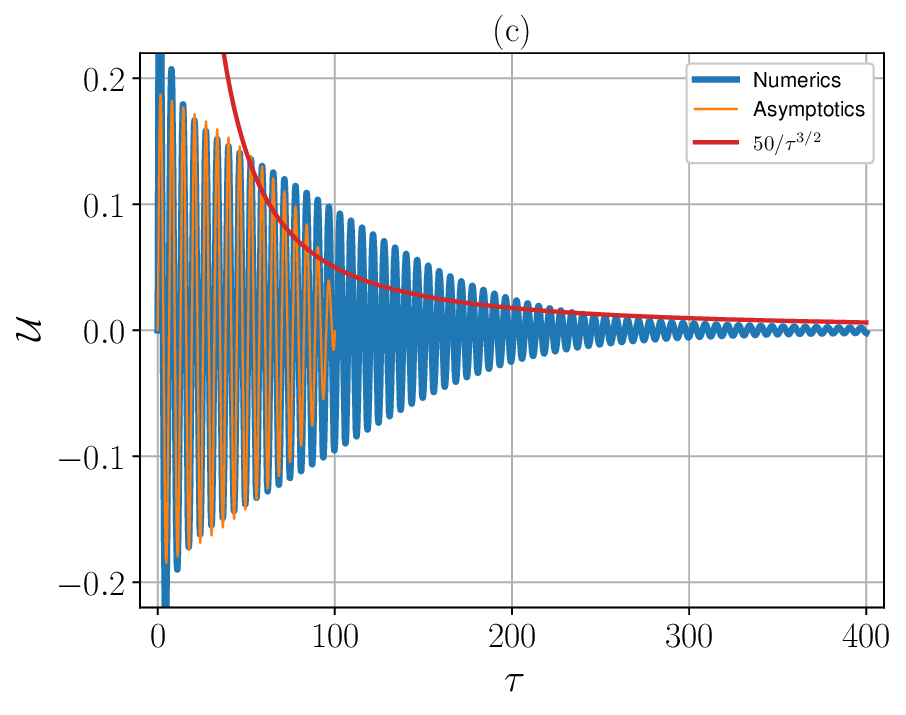}
\caption{(a) Given time-varying parameter $K$, (b) the trapped
mode frequency $\Omega_0$, (c) the displacement $\mathcal U$ of the discrete sub-system 
for the case when parameters
are taken according Eqs.~\eqref{2p-asT},
\eqref{K-exp}}
\label{all3-time.pdf}
\end{figure}
\afterpage{\clearpage}

\section{Conclusion}

In the paper, we have obtained the leading-order term of a universal
asymptotics, which can approximate the free non-stationary response $\mathcal U$ of a discrete
mass-spring-damper system embedded into a continuum system governed
by the telegraph PDE. All parameters of such a composite system are
assumed to be slowly time-varying functions. Additionally, the parameters of
the continuum sub-system can be spatially varying functions. 
In a particular case, the discrete sub-system can move along the
continuum one at a sub-critical speed, which is also
assumed to be a slowly time-varying function. 
We propose the solution in the form of the single-mode ansatz
\eqref{ansatz-u}--\eqref{ansatz-u-ext},
which corresponds to the dominant frequency in the non-stationary
response of the corresponding system with constant coefficients, i.e., the
trapped mode frequency
\eqref{FR_0}.
We require that the localization conditions, see Remark
\eqref{loc-remark},
are always fulfilled for the instantaneous values of the system parameters in a certain
neighbourhood of the discrete sub-system.
The most important result of the
paper is formula
\eqref{final-evolution} for the leading-order term $\WF_0(T)$ of the
non-stationary oscillation
amplitude. The quantities involved into Eq.~\eqref{final-evolution} 
 should be found by Eqs.~\eqref{WW_2},
\eqref{Co-value}.
The asymptotics 
for the response $\mathcal U$ in the real form is given by Eq.~\eqref{U-expr}.

Though mathematical technique used in the current paper and suggested
first time in \cite{gavrilov2002etm} is inspirited by the formal procedure of obtaining
the Liouville-Green (or WKB) approximation by the method of multiple scales \cite{nayfeh2008perturbation}, 
the asymptotic ansatz that we use is quite similar to one used in
the space-time ray method \cite{Babich2009,Babich2002}. 
The essential difference from the first approach is that we deal with a PDE coupled
with an ODE, whereas the Liouville-Green approximation was suggested for an
ODE. Comparing our approach with the space-time ray method, one can see that
we deal with a pure non-stationary solution. Even in zeroth order
approximation, the non-stationary solution is a Fourier integral over all
possible frequencies. Therefore, we 
need an additional step to satisfy the initial conditions,
where we match (see Remark~\ref{remark-matching})
the single-mode asymptotics 
with the results obtained by 
the method of stationary phase applied to the same system with
constant parameters, which are equal to the corresponding initial values. 
{In the framework of the space-time ray method the initial conditions are
usually formulated in terms of quantities involved in the ansatz (the
amplitude and the phase), therefore the matching is not required.}
Another important difference of our problem, compared with problems where the space-time
ray method is usually used, is that the equations for the amplitude (the transport
equations in the framework of the space-time ray method) in our case can be equivalently 
reduced, at least in the first approximation, to not a PDE, but to a single ODE.


According to Eqs.~\eqref{WW_2}--\eqref{WW_final}, in the non-dissipative case
\eqref{nondiss-ass},
the leading-order term of the expansion for the amplitude of
oscillation is obtained {in the form of an algebraic expression, 
which involves the instantaneous limiting values of the system
parameters independently varying accordingly to unknown arbitrary laws.}         
In the dissipative case
\eqref{diss-ass},
we, generally, obtain the leading-order term of the expansion for the
amplitude {in quadratures} as a functional, which depends on the history
of the system parameters, though in some exceptional cases the result can be
obtained as a function of time and the instantaneous limiting values of the system
parameters.
From the physical point of view,
the obtained results are an essential extension, related to 
the localized oscillation,
of the classical result 
\eqref{App1-sol1}
concerning the oscillation of a mass-spring-damper system with 
time-varying parameters,
see Appendix~\ref{App-1d}.

By construction, the first approximation equation \eqref{WW} 
has the structure of Eq.~\eqref{gen-K-M-1st}, 
where $\pp$ and $\tilde\pp$ are system parameters introduced by 
Eqs.~\eqref{pP}, \eqref{pPP}. The most general form of the solution of 
Eq.~\eqref{gen-K-M-1st} is functional 
\eqref{q2}, which depends on the history of the system parameters.
Nevertheless, if 
the right-hand side of Eq.~\eqref{gen-K-M-1st} (or Eq.~\eqref{WW}) can be represented 
in the form of the exact derivative of a
certain function of the system parameters,
see Eq.~\eqref{exact-d-P}, then the corresponding solution is a
function, which depends on the instantaneous limiting values of the system
parameters, see Eq.~\eqref{q11}. Such a representation is possible if and only
if conditions \eqref{PD1} and \eqref{PD2} are fulfilled. Note that 
Eq.~\eqref{PD1} is always true for any non-dissipative system under
consideration.
For non-dissipative 
problems with a single time-varying parameter, considered in our
previous studies 
\cite{Gavrilov2022jsv,Gavrilov2019nody,Gavrilov2019asm,indeitsev2016evolution,gavrilov2002etm},
condition \eqref{PD2}
is also always true, and the solution of the first approximation equation can be
found in the form of certain functions, see Remark~\ref{remark-one-parametric}.
In the non-dissipative case of several time-varying parameters considered in this paper, the possibility to represent a given equation with the
structure of Eq.~\eqref{gen-K-M-1st} in the form of 
Eq.~\eqref{exact-d-P} is absolutely unobvious. We can try to check
conditions 
\eqref{PD2} directly. However, this can be a difficult problem due to a lengthy structure of
the right-hand side of the equation under consideration. 
On the other hand, it is possible to introduce new variables $\bPsi(\pp)$ instead
of $\pp$ and equivalently rewrite the first approximation equation in the form of 
Eq.~\eqref{gen-K-M-1st-psi}. 
A lucky choice of the variables $\bPsi(\pp)$ can essentially 
simplify the
calculations and help to obtain the representation of Eq.~\eqref{gen-K-M-1st} in the
form of Eq.~\eqref{exact-d}.
This trick was suggested by
Poroshin in \cite{Gavrilov2022jsv} and the same approach is applied in this
paper.
The existence of such a representation immediately guaranties us that the
amplitude can be obtained as a function, not as a functional, and gives us the
possibility to calculate this function in an explicit form. 
Generally, it is not 
clear for us, due to which properties of governing equations \eqref{eq2m},
\eqref{eq1m} with
variable coefficients, we have obtained 
the first approximation equation for the
amplitude $\WF_0$ with the exact derivative in the right-hand side.
Note that magically the exact derivatives can
be met often when dealing with WKB approximations.  For example, the
right-hand sides of the odd successive WKB approximations for ODE describing
a mass-spring system with time-varying stiffness, i.e., for
Eq.~\eqref{App1-baseq-f} wherein $M=\text{const}$ and $\varGamma=0$,
are the exact derivatives \cite{Sukumar2023}, see Remark~\ref{remark-sro}.
Thus, the possibility to obtain the result 
for the amplitude in a closed form 
for system with several simultaneously
varying parameters is not related with the summarization of the
obtained in previous papers 
\cite{Gavrilov2019nody,Gavrilov2019asm,indeitsev2016evolution,gavrilov2002etm}
results, but with the trick suggested in the recent
paper \cite{Gavrilov2022jsv},
see Sect.~\ref{sect-disc}.

The practical applicability of the obtained in such a way formal single-mode
asymptotics should be verified
numerically.  This has been done in our previous papers 
\cite{Gavrilov2022jsv,Gavrilov2019nody,Gavrilov2019asm,indeitsev2016evolution,gavrilov2002etm}
for a number of
single-parameter perturbed non-dissipative systems. In the current paper, 
a two-parameter perturbed system was considered, see Sect.~\ref{Sect-two-pa}, as well as a
dissipative one, see Sect.~\ref{Sect-num-diss}. In all cases, an excellent agreement was
demonstrated. The fact that the single-mode approximation is in an excellent
agreement with numerics, is related, in our opinion, with the phenomenon of
the anti-localization of non-stationary linear waves, discovered recently
\cite{Gavrilov2024,Shishkina2023cmat,Shishkina2023jsv,Gavrilov2023DD}, see Sect.~\ref{As-breaks}.


One of the practically important problems, 
for which the asymptotics obtained in this paper appears to be
applicable,
is the problem concerning 
a mass-spring-damper system 
non-uniformly moving along the wave-guide with the spatially varying stiffness
$k$. Since this is a non-symmetrical with respect to $\xi$ problem, for now, we are
not ready to present the corresponding numerical calculations,
see Remark~\ref{remark-sym}. On the other hand, this problem has many
specific important for applications peculiarities,  which can make
it the subject for a separate study.

\section*{Acknowledgement}
The authors are grateful to A.P.~Kiselev, A.M.~Krivtsov, O.V.~Motygin,
M.V.~Perel for discussions. S.N.G.\ is grateful to Prof. D.A.~Indeitsev, 
who many years ago had told him that  
a string on the Winkler foundation with
a moving inertial inclusion is a system with a trapped mode and thereby had initiated 
this study.
This paper is dedicated to the memory of Prof.~D.A.~Indeitsev.
The research is supported by the Ministry of Science and Higher Education of
the Russian Federation (project 124040800009-8).

\appendix

\section{Oscillation of a mass-spring-damper system with time-varying parameters}
\label{App-1d}
Consider, {mostly following the formal procedure used by Nayfeh} 
(Sect.~7.1.6 in \cite{nayfeh2008perturbation}),
a free non-stationary oscillation of a mass-spring-damper system.  The
viscosity of the dumper is a small quantity $\epsilon
\varGamma$. 
\begin{remark}	
Nayfeh did not use the last assumption. We use it, to make the problem similar to one
formulated in Sect.~\ref{Sect-formulation}, see also  
Remark~\ref{remark-small-diss}.
\end{remark}
The stiffness $K(\epsilon \t)$,
the mass 
$M(\epsilon \t)$,  and the viscosity $\epsilon
\varGamma(\epsilon \t)$ are slowly time-varying quantities. 
The governing
equation is
\begin{equation}
( M(\epsilon \t)  \,\dot{\mathcal U})^\bcdot
+\left\{2\epsilon\GA(\epsilon \t)\, \dot{\mathcal U}\right\} + K(\epsilon \t)\,
{\mathcal U} = 0,
\label{App1-baseq}
\end{equation}
where $\mathcal U$ is the displacement. 
\begin{remark}	
Generally, Nayfeh (\cite{nayfeh2008perturbation}) assumed  that coefficients $M$,
$\varGamma$, and $K$ are given in the form of
asymptotic power series in $\epsilon$:
\begin{equation}
M=\sum_n M_n(\epsilon \t)\epsilon ^n,
\qquad
\GA=\sum_n \GA_n(\epsilon \t)\epsilon ^n,
\qquad
K=\sum_n K_n(\epsilon \t)\epsilon ^n.
\label{coeff-ex}
\end{equation}
Here, we assume that {the only} non-zero terms 
in expansions \eqref{coeff-ex} correspond to $M_0\equiv M$, $K_0\equiv K$,
and $\GA_0\equiv\GA$.
In 
\eqref{App1-baseq} 
the only term in the left-hand side, which 
is proportional to $\epsilon$ is shown in
the curly brackets.
\end{remark}
In the same way as we have done it, see 
Eqs.~\eqref{pP},
\eqref{pPP},
for the full system, we can characterize the properties of the system described by 
\eqref{App1-baseq} by $n$-tuple ($n=2$) 
\begin{equation}
\pp(\epsilon\tau)\=(\p_1,\p_2)\in\mathbb R^2,
\qquad
\p_1=K,
\quad 
\p_2=M,
\label{pP-A}
\end{equation}
of parameters for the zero-order system, and an additional $\tilde n$-tuple 
($\tilde n=1$) 
\begin{equation}
\ppp(\epsilon\tau)\=(\tilde\p_1)\in\mathbb R^1,\qquad
\tilde\p_1=\varGamma,
\label{pPP-A}
\end{equation}
of parameters, which are necessary to consider additionally to $\pp$ for the system with $\epsilon>0$.

We introduce {the} slow time $T=\epsilon t$ and 
represent the solution in the form
of the following ansatz 
%
\begin{gather}
\mathcal U=\mathcal W\,\exp \phi+\cc,
\label{App1-U-phi}
\end{gather}
where 
\begin{gather}
\mathcal W(T)=\sum_i \epsilon^i\mathcal W_i(T),
\\
\dot\phi(T)=-\I\Omega_0(T)
\end{gather}
are the amplitude and the phase, respectively;
\begin{equation}
\Omega_0(T)=\sqrt{\frac{K(T)}{M(T)}}
\label{App1-Omega0}
\end{equation}
is the natural frequency. Then, we consider 
$T$
and $\phi$ as independent time-like variables, and use the corresponding representations 
\eqref{dot-xt} for differential operators with respect to time.
Substituting Eqs.~\eqref{App1-U-phi}--\eqref{App1-Omega0} into 
Eq.~\eqref{App1-baseq} results in the following first approximation equation for 
the leading-order term $\mathcal W_0(T)$:
\begin{equation}
2\sqrt{MK}\,{\mathcal W}_0{}'_T+M\left(\sqrt{\frac{K(T)}{M(T)}}\,\right)'_T\mathcal W_0+{M}{}'_T\sqrt{\frac{K(T)}{M(T)}} \mathcal W_0+
\left\{
2\varGamma\sqrt{\frac{K(T)}{M(T)}} \mathcal W_0
\right\}
=0,
\label{long-P}
\end{equation}
which is written in terms of variables $\pp$, $\tilde \pp$. Here and in what
follows in Appendix~\ref{App-1d}, the contribution from the term in the
left-hand side of 
\eqref{App1-baseq}, which is proportional to $\epsilon$, is again shown in the
curly brackets.
Taking into account that 
\begin{equation}
\left(\sqrt{\frac{K(T)}{M(T)}}\,\right)'_T
=
\frac{K'_T}{2\sqrt{KM}}-\frac{M'_T\sqrt K}{2M^{3/2}},
\end{equation}
after simplification, one can rewrite 
Eq.~\eqref{long-P} in the following form:
\begin{equation}
\frac{\mathcal W_0{}'_T}{\mathcal W_0}=
-\frac{K'_T}{4K}-\frac{M'_T}{4M}-
\left\{
\frac{\varGamma}{M}
\right\}
.
\label{App1-1st-P}
\end{equation}
The last equation has the structure of Eq.~\eqref{gen-K-M-1st}.
Provided that 
$\varGamma=0$,
we get the solution of 
Eq.~\eqref{gen-K-M-1st}
in the form of a function, which depends on the
instantaneous values of the system parameters $\pp$ only:
\begin{equation}
  \WF_{0}(T)
  =\frac {\mathcal C}
  {\sqrt[4]{M(T) K(T)}}
  = \frac 
  {\mathcal C}
  {\sqrt{M(T)\Omega_0(T)}}.
\label{App1-sol-0}
\end{equation}
In the particular case $\dot M=0$, the amplitude $\mathcal W_0$
is proportional to the inverse of the square root of the natural frequency.
\begin{equation}
\mathcal W_0(T)=\frac {\bar {\mathcal C}}{\sqrt{\Omega_0(T)}},
\label{L-G}
\end{equation}
where $\bar{\mathcal C}=\mathcal C/\sqrt M$ is an arbitrary constant.
This formula now is well-known as the 
{Liouville -- Green} or {WKB} 
approximation \cite{nayfeh2008perturbation}, see also
\cite{feschenko1967eng,McHugh1971} for historical aspects. If
$\varGamma=\const\neq0$ and $M=\const$, we again 
obtain the solution in the form of a function, which depend on current values of
the system parameters $\pp$, and, in the latter case, the current value of
time~$T$:
\begin{equation}
  \WF_{0}(T) 
  = \frac {\mathcal C\Exp{
  -\frac{\varGamma T}M
  }}
  {\sqrt{M\Omega_0(T)}}
  = \frac {\bar{\mathcal C}\Exp{
  -\frac{\varGamma T}M
  }}
  {\sqrt{\Omega_0(T)}}.
\label{App1-sol}
\end{equation}
Finally, if any of
$\varGamma$ or $M$ is not a constant,
one obtains the solution in quadratures in the form of a functional: 
\begin{equation}
  \WF_{0}(T) 
  =\frac {\mathcal C\Exp{-\int_0^T \frac {\varGamma(\tp) \,\d  \tp}{M(\tp)}}}
  {\sqrt[4]{M(T) K(T)}}
  = \frac {\mathcal C\Exp{-\int_0^T \frac {\varGamma(\tp)\,\d \tp} {M(\tp)} }}
  {\sqrt{M(T)\Omega_0(T)}}
  .
\label{App1-sol1}
\end{equation}

\begin{remark}	
The right-hand side of 
\eqref{App1-1st-P} satisfies condition \eqref{PD2} since
\begin{equation}
\pd{F_0}{\p_1}=\pd{\left(-\frac1{4\p_0}\right)}{\p_1}=0,
\qquad
\pd{F_1}{\p_0}=\pd{\left(-\frac1{4\p_1}\right)}{\p_0}=0.
\end{equation}
\end{remark}

Note that for the case of
Eq.~\eqref{App1-baseq}, instead of $\pp$, it is useful to take new variables:
\begin{equation}
\Psi_1(T)=\Omega_0(T), 
\qquad
\Psi_2(T)=M(T).
\label{psi-omega}
\end{equation}
Instead of Eq.~\eqref{long-P} we get
\begin{equation}
2M\Omega_0{\mathcal W}_0{}'_T+M{\Omega_0}{}'_T\mathcal W_0+{M}{}'_T\Omega_0 \mathcal W_0+
\left\{
2\varGamma\Omega_0 \mathcal W_0
\right\}
=0
\end{equation}
or
\begin{equation}
\frac{\mathcal W_0{}'_T}{\mathcal W_0}=
-\frac{\Omega_0{}'_T}{2\Omega_0}-\frac{M'_T}{2 M}-
\left\{
\frac{\varGamma}{M}
\right\}
.
\label{App1-1st}
\end{equation}
Provided that $\varGamma=0$, 
Eq.~\eqref{App1-1st} has the structure of Eq.~\eqref{exact-d},
and the solution is given by 
Eq.~\eqref{App1-sol-0}. The solution in case $\varGamma\neq0$ can be obtained 
analogously, in the same way, as it was done 
when deriving formulae \eqref{App1-sol}--\eqref{App1-sol1}. One can see that
in terms of variables $\bPsi$
defined by Eq.~\eqref{psi-omega} the calculations are essentially simpler than for
the original parameters~$\pp$.

\begin{remark}	
\label{remark-sro}
Equation 
\eqref{App1-baseq}
with time-varying coefficients $K$ and $\varGamma$ is considered in Nayfeh
book \cite{nayfeh2008perturbation}. In  
\cite{feschenko1967eng} equation 
\begin{equation}
 M(\epsilon \t)  \,\ddot{\mathcal U}
+2\GA(\epsilon \t)\, \dot{\mathcal U} + K(\epsilon \t)\,
{\mathcal U} = 0
\label{App1-baseq-f}
\end{equation}
is discussed instead of Eq.~\eqref{App1-baseq}. This equation can also be reduced 
to the form of Eq.~\eqref{App1-baseq}
with only time-varying coefficients $K$ and $\varGamma$.
Equation 
\eqref{App1-baseq} with $\varGamma=0$ is discussed in book
\cite{Kevorkian1996} (Sect.~4.3.3), where solution
\eqref{App1-sol-0} is found. {In all these studies, the question if the
expression for the leading-order term $\WF_0$ of the amplitude of the solution of
a certain ODE with variable coefficients can be represented as a function, depending on the current
values of the system parameters only, or as a functional, is not discussed.
Moreover, we did not manage to find any studies where this problem was noticed,
although there may be some. 
Note that, the
right-hand sides of successive odd WKB approximations for 
Eq.~\eqref{App1-baseq} wherein $M=\text{const}$ and $\varGamma=0$
are the exact derivatives, see \cite{Sukumar2023}, where such an equation
treated as the one-dimensional time-independent Schr\"odinger equation with a spatial
variable~$\tau$.   
}
\end{remark}

%
%
%


\section{The properties of the zeroth order system}
\label{App-e0}

\label{app-dispersion}
In Appendix~\ref{App-e0}, we provide the necessary formulae for the system
under consideration with $\epsilon=0$, i.e., for the composite system with
constant parameters $\pp$ introduced by
Eq.~\eqref{pP}
and zero dissipation, see Eq.~\eqref{nondiss-ass}. Accordingly,
Eqs.~\eqref{eq2m} and \eqref{eq1m} can be rewritten as
\begin{gather}
\label{eq2const}
 M \ddot\UF
 + K \UF = -P(\tau) +p(\tau),
\\
\label{eq1const}
(\TF -\rho\v{^2}) u''+ 2 v\rho {\dot u'}  - \rho\ddot u  - k u =
-P(\tau)\delta(\xi).
\end{gather} 
All results are obtained in the dimensionless form in 
\cite{Gavrilov2022jsv}. Some particular results were obtained in
\cite{Gavrilov2019asm,kaplunov1986torsional,kaplunov1995simple,Kruse1998,Glushkov2011a}.

\subsection{The dispersion relation}
Assuming that $p(\tau)=0$ and 
\begin{equation}
\label{ansatz_dis}
u=\WF\e^{-\I(\O \t - \omega \xi)}
\end{equation}
in intervals $\xi\gtrless0$,
we get the dispersion relation for the operator in the left-hand side of Eq.~\eqref{eq1const}:
\begin{equation}
\label{dis_rel}
\omega^2-2B(\O)\omega+A^2(\O) =0.
\end{equation}
Here $\omega$ is the wave-number:
\begin{gather}
\label{omega}
\omega = B(\O) \pm  \I S(\O),
\\
\label{B}
B(\O)=\frac{\v\O\rho}{\TF- \rho\v^2} 
\\
\label{A}
A^2(\O)=\frac{k-\rho \O^2}{\TF- \rho\v^2},
\\
\label{S}
S(\O) = \sqrt{A^2(\O) - B^2(\O)} = \frac{\sqrt{k\TF-k\rho\v^2-\TF\rho\O^2}}{\TF-\rho\v^2} 
=\frac{k}{\sqrt{\rho \TF}} \frac {\sqrt{\O^2_*-\O^2 }}{\O^2_*}.
\end{gather}
The quantity $\O_\ast$ is the cut-off (boundary) frequency
\begin{equation}
\label{cutoff}
\O_\ast = \frac{k}{\rho}-\frac{k v^2}{\TF}=\frac{k}{\rho \TF}(\TF-\rho v^2),
\end{equation}
which separates the pass-band
$\Omega^2>\Omega_\ast^2$, where the
solution is a sinusoidal wave and the stop-band
$0<\Omega^2<\Omega_\ast^2$
, where the solution is an
inhomogeneous wave. 
\begin{remark}	
Note that according Eqs.~\eqref{restr-par-21},
\eqref{v(t)and_c},
\eqref{c-def} $\Omega_\ast>0$.
\end{remark}

\subsection{Spectral problem for a trapped mode}
\label{App-trapped}
Put $p=0$ and consider the spectral problem concerning free oscillation
of the system described by Eqs.~\eqref{eq2const}, \eqref{eq1const}. We consider
only sub-critical speeds, i.e., Eq.~\eqref{v(t)and_c} is fulfilled.
Let
\begin{gather}
\label{ansatz_sp1}
u=W(\xi)\,\e^{-\I\O \t}, \\
\label{ansatz_sp2}
\UF =\WF\, \e^{-\I\O \t}.
\end{gather}
Trapped modes are modes with finite energy, i.e., we require
\begin{equation}
\int_{-\infty}^{+\infty}W^2\,\d{\xi}<\infty,\quad \int_{-\infty}^{+\infty}W'^2\,\d{\xi} < \infty.
\end{equation}

By substituting Eqs.~\eqref{ansatz_sp1}, \eqref{ansatz_sp2} into
Eqs.~\eqref{eq2const}, \eqref{eq1const}, one gets
\begin{equation}
\label{spect}
(\TF- \rho\v^2)\big(W''-2\I B(\O) W' - A^2(\O) W\big) =
-  (M \O^2 - K) \WF\,  \delta (\xi).
\end{equation}
The solution of Eq.~\eqref{spect} can be written as follows:
\begin{equation}
\label{W}
W(\xi) = (M \O^2 - K) \WF G(\xi,\O),
\end{equation}
where $ G(\xi,\O)$ is the Green function in the frequency domain. For 
$\Omega^2 < \Omega_\ast^2$ one has
\begin{equation}
G(\xi,\O)= \frac{\exp{(\I B \xi - S|\xi|)}}{2(\TF-\rho\v^2)S}.
\end{equation}
Taking into account Eqs.~\eqref{B}, \eqref{S}, one gets:
\begin{equation}
\label{WG}
W(\xi) = \frac{(M \O^2 - K) \WF}{2\sqrt{k\TF-k\rho\v^2-\TF\rho\O^2}}\,
\text{exp}\left(\frac{\I\v\O\rho\xi- \sqrt{k\TF-k\rho\v^2-\TF\rho\O^2}|\xi|}{\TF-\rho\v^2}\right).
\end{equation}
Calculating Eq.~\eqref{WG} at $\xi=0$ yields the frequency equation for the
trapped mode frequency $\O_0$:
\begin{equation}
\label{fr_eq}
2\sqrt{k\TF-k\rho\v^2-\TF\rho\O_0^2}=M\O_0^2-K.
\end{equation}
The last equation can be equivalently rewritten in the following form:
\begin{equation}
2\sqrt{\O_\ast^2-\O_0^2}=\frac{M\O_0^2-K}{\sqrt{\TF \rho}}.
\label{fr_eq-1}
\end{equation}
The positive roots $\Omega_0^2>0$ of the frequency equation 
\eqref{fr_eq-1} are trapped modes frequencies. It can be shown that provided
that Eqs.~\eqref{restr-par-11}, \eqref{restr-par-21},
\eqref{v(t)and_c} 
as well as the following inequalities 
\begin{gather}
-2\sqrt{\TF \rho }|\Omega_\ast|<K,
\label{stab-cond}
\\
K<M\Omega^2_\ast 
\label{loc-cond}
\end{gather}
are true, there exists a unique trapped mode with the frequency
\begin{equation}
\label{FR_0}
\begin{aligned}
&\O^2_0=\frac{KM-2\TF\rho+2\sqrt{\TF\rho(\TF\rho-KM)+M^2k(\TF-\rho\v^2)
}}{M^2},&&\qquad M>0;
\\
&\O^2_0 = \O^2_* - \frac{K^2}{4\TF\rho},
&&\qquad M=0.
\end{aligned}
\end{equation}
Here $\Omega_\ast$ is defined by Eq.~\eqref{cutoff}.
\begin{remark}  
The root
\eqref{FR_0}
of frequency equation \eqref{fr_eq} satisfies inequalities
\begin{equation}
\begin{aligned}
\label{interval}
\frac{K}{M} &< \O^2_0 < \O_*^2,\quad &M>0\ \text{and}\ K\geq 0,
\\
0 &< \O^2_0 < \O_*^2,\quad & K<0.
\end{aligned}
\end{equation}
\end{remark}

\begin{remark}	
\label{loc-remark}
The domain defined by
Eqs.~\eqref{restr-par-11}, \eqref{restr-par-21},
\eqref{v(t)and_c},
\eqref{stab-cond},
\eqref{loc-cond}
in the six-dimensional parameter space with
co-ordinates 
\eqref{pP}
is the localization domain $\LL$. Condition \eqref{stab-cond} has the meaning of the
stability condition, which is important for $K\leq0$. Violating
Eq.~\eqref{stab-cond} leads to zeroing of the trapped mode frequency
$\Omega_0=0$ and 
a localized buckling of the continuum
sub-system. 
Approaching of the system parameters to the boundary 
\eqref{loc-cond} corresponds to approaching the trapped mode frequency
$\Omega_0$
to the cut-off frequency $\Omega_\ast$:
\begin{equation}
\lim_{K\to M\Omega_\ast-0} \Omega_0=\Omega_\ast
\end{equation}
i.e., to the upper boundary of the
stop-band, and, finally, to the disappearing of the trapped mode. 
\end{remark}
\begin{remark}	
Conditions 
\eqref{restr-par-11}, \eqref{restr-par-21} are some pre-assumed physical restrictions,
which are not directly related to the existence of the trapped mode. Moreover,
as shown in \cite{Glushkov2011a}
for $M<0$ even two trapped modes can exist. Nevertheless, we do not know any
physical realization for a system with $M<0$ (though there may be some) and do
not take into account such a case.
\end{remark}

\begin{remark}	
It can be shown that the expression for $\Omega_0^2$ defined by 
\eqref{FR_0} is continuous at $M=+0$.
\end{remark}

\begin{remark}	
The real solution of Eq.~\eqref{eq2const} wherein $p=0$ and Eq.~\eqref{eq1const}, which correspond
to the trapped mode, is
\begin{equation}
u=C\Exp{-S(\Omega_0)|\xi|}\cos(\Omega_0\t-B(\Omega_0)\xi-D),
\label{eq:solution_inhomogeneous_stationary_problem-ext}
\end{equation}
where $C$, $D$ are arbitrary real constants.

\end{remark}

\subsection{Non-stationary free oscillation}
\label{app-non-st}

Consider now the non-stationary problem with the external loading $p(\tau)$
satisfying 
Eq.~\eqref{pulse}. The solution can be obtained in the form of the
(generalized) inverse
Fourier transform

\begin{equation}
\label{F_inv}
\UF(\t) = \frac{1}{2\pi}\int\limits_{-\infty}^{+\infty} \frac{\F{p}(\Omega)\,
\e^{-\I\O \t}\d\O}{ 2\sqrt{k\TF-k\rho\v^2-\TF\rho\O^2} + K-M\O^2},
\end{equation}
where symbol $\F{p}(\Omega)$ denotes the value of the Fourier transform for the
loading $p(\t)$ calculated at a frequency $\Omega$. The large time asymptotics for the right-hand
side of Eq.~\eqref{F_inv} can be found by the method of stationary phase 
\cite{Fedoryuk1977} and has the following form \cite{Gavrilov2022jsv}:
\begin{equation}
\UF(\t) =
\frac{
    \sqrt{k\TF-k\rho\v^2-\TF\rho\O_0^2}\,|\F{p}(\O_0)|
}{\O_0\left(\TF\rho+M \sqrt{kT-k\rho\v^2-\TF\rho\O_0^2}\right)}
\sin\big(\O_0 \t - \text{arg}(\F{p}(\O_0))\big)+\text{o}(1), ~~\t{\to\infty}.
\label{eq:solution_inhomogeneous_stationary_problem}
\end{equation}
Hence, for the large times, the oscillation with 
the trapped mode frequency $\O_0$ is dominant and only non-vanishing
component of the
non-stationary response for the system under
consideration.

\subsection{Derivation of identity~\eqref{relationship}}
\label{App-i}
According to Eq.~\eqref{S}, we have
\begin{equation}
S'_{\O}=\frac{-\TF \rho \Omega}{ (\TF-\rho v^2)\sqrt{k\TF-k \rho v^2 -\TF\rho
\Omega^2} }
\end{equation}
and, therefore,
\begin{equation}
(\TF - \rho \v^2) S'_{\O}-M\O=-\frac{\TF \rho \Omega+M \Omega\sqrt{k\TF-k \rho v^2 -\TF\rho
\Omega^2} }{\sqrt{k\TF-k \rho v^2 -\TF\rho
\Omega^2}}.
\label{der-1}
\end{equation}
Due to Eqs.~\eqref{B}, \eqref{S} we get:
\begin{equation}
\rho\v B +\rho \O + M\O S=\frac{\TF \rho \Omega + M \Omega \sqrt{k\TF
-k\rho v^2 -\TF \rho \Omega^2}}{\TF- \rho v^2}.
\label{der-2}
\end{equation}
Substituting Eqs.~\eqref{der-1}, \eqref{der-2} into the left-hand side of 
Eq.~\eqref{relationship},
and using Eq.~\eqref{S}, we, finally, obtain:
\begin{equation}
 \frac{	\left( \TF - \rho \v^2 \right) S'_{\O}-	M\O}
{\rho\v B +\rho \O + M\O S } = -\frac{\TF-\rho v^2}{\sqrt{k\TF-k \rho v^2 -\TF\rho
\Omega^2}}= -\frac{1}{S}.
\end{equation}

\bibliographystyle{plainnat}
\bibliography{bib/serge-gost,bib/all}


\end{document}